\documentclass[longauth]{aa}
\usepackage[varg]{txfonts}
\usepackage{natbib}
\usepackage{lipsum}
\usepackage{subfigure}
\usepackage{adjustbox}
\usepackage{graphicx}
\usepackage{multirow}
\usepackage{xcolor}
\usepackage[colorlinks, citecolor=blue, linkcolor=black]{hyperref}
\bibpunct{(}{)}{;}{a}{}{,}

\usepackage[normalem]{ulem}

\begin{document}

\title{MOLsphere and pulsations of the Galactic Center's red supergiant GCIRS 7 from VLTI/GRAVITY{\thanks{Based on observations made with ESO Telescopes at the La Silla Paranal Observatory under the programme IDs 098.D-0250 and 103.B-0032.}}}

\author{GRAVITY Collaboration\thanks{GRAVITY is developed in a collaboration by the Max Planck Institute for extraterrestrial Physics, LESIA of Observatoire de Paris / Université PSL / CNRS / Sorbonne Université / Université de Paris and IPAG of Université Grenoble Alpes / CNRS, the Max Planck Institute for Astronomy, the University of Cologne, the CENTRA - Centro de Astrofisica e Gravitação, and the European Southern Observatory. Corresponding author: G.~Rodríguez-Coira.}: G. Rodríguez-Coira \inst{\ref{ins1}}
\and T.~Paumard \inst{\ref{ins1}}
\and G.~Perrin \inst{\ref{ins1}}
\and F.~Vincent \inst{\ref{ins1}}
\and R.~Abuter \inst{\ref{ins8}}
\and A.~Amorim \inst{\ref{ins6},\ref{ins13}}
\and M.~Bauböck \inst{\ref{ins2}}
\and J.~P.~Berger\inst{\ref{ins5}}
\and H.~Bonnet \inst{\ref{ins8}}
\and W.~Brandner \inst{\ref{ins3}}
\and Y.~Clénet \inst{\ref{ins1}}
\and P.~T.~de~Zeeuw \inst{\ref{ins11},\ref{ins2}}
\and J.~Dexter \inst{\ref{ins2}, \ref{ins14}}
\and A.~Drescher \inst{\ref{ins2}}
\and A.~Eckart \inst{\ref{ins4}, \ref{ins10}}
\and F.~Eisenhauer \inst{\ref{ins2}}
\and N.~M.~Förster Schreiber \inst{\ref{ins2}}
\and F.~Gao \inst{\ref{ins2}}
\and P.~Garcia \inst{\ref{ins7},\ref{ins13}}
\and E.~Gendron \inst{\ref{ins1}}
\and R.~Genzel \inst{\ref{ins2},\ref{ins12}}
\and S.~Gillessen \inst{\ref{ins2}}
\and M.~Habibi \inst{\ref{ins2}}
\and X.~Haubois \inst{\ref{ins9}}
\and T.~Henning \inst{\ref{ins3}}
\and S.~Hippler \inst{\ref{ins3}}
\and M.~Horrobin \inst{\ref{ins4}}
\and A.~Jimenez-Rosales \inst{\ref{ins2}}
\and L.~Jochum \inst{\ref{ins9}}
\and L.~Jocou \inst{\ref{ins5}}
\and A.~Kaufer \inst{\ref{ins9}}
\and P.~Kervella \inst{\ref{ins1}}
\and S.~Lacour \inst{\ref{ins1}}
\and V.~Lapeyrère \inst{\ref{ins1}}
\and J.~B.~Le~Bouquin \inst{\ref{ins5}}
\and P.~Léna \inst{\ref{ins1}}
\and M.~Nowak \inst{\ref{ins17},\ref{ins1}}
\and T.~Ott \inst{\ref{ins2}}
\and K.~Perraut \inst{\ref{ins5}}
\and O.~Pfuhl \inst{\ref{ins8}}
\and J.~Sanchez-Bermudez \inst{\ref{ins3}, \ref{ins18}}
\and J.~Shangguan \inst{\ref{ins2}}
\and S.~Scheithauer \inst{\ref{ins3}}
\and J.~Stadler \inst{\ref{ins2}}
\and O.~Straub, \inst{\ref{ins2}}
\and C.~Straubmeier \inst{\ref{ins4}}
\and E.~Sturm \inst{\ref{ins2}}
\and L.~J.~Tacconi \inst{\ref{ins2}} 
\and T.~Shimizu \inst{\ref{ins2}}
\and S.~von~Fellenberg\inst{\ref{ins2}}
\and I.~Waisberg \inst{\ref{ins16},\ref{ins2}}
\and F.~Widmann \inst{\ref{ins2}}
\and E.~Wieprecht \inst{\ref{ins2}}
\and E.~Wiezorrek \inst{\ref{ins2}}
\and J.~Woillez \inst{\ref{ins8}}
\and S.~Yazici \inst{\ref{ins2},\ref{ins4}}
\and G.~Zins \inst{\ref{ins9}}
}

\authorrunning{GRAVITY Collaboration}
\titlerunning{MOLsphere and pulsations of GCIRS 7 from VLTI/GRAVITY}

\institute{Max Planck Institut fur Exterterrestrische Physik (MPE), Giessenbachstr.1, D-85748 Garching, Germany \label{ins2}
\and LESIA, Observatoire de Paris, Universite PSL, CNRS, Sorbonne Université, Université de Paris, 5 place Jules Janssen, F-92195 Meudon, France \label{ins1}
\and Max Planck Institute for Astronomy, Königstuhl 17, 69117 Heidelberg, Germany \label{ins3}
\and 1. Physikalisches Institut, Universitat zu Köln, Zülpicher Stra{\ss}e 77, 50937 Cologne, Germany \label{ins4}
\and CNRS, IPAG, Univ.Grenoble Alpes, F-38000, Grenoble, France \label{ins5}
\and Universidade de Lisboa - Faculdade de Ciências, Campo Grande, P-1749-016 Lisboa, Portugal \label{ins6}
\and Faculdade de Engenharia, Universidade do Porto, Rua Dr. Roberto Frias, P-4200-465 Porto, Portugal \label{ins7}
\and European Southern Observatory, Karl-Schwarzschild-Stra{\ss}e 2, 85748 Garching, Germany \label{ins8}
\and European Southern Observatory, Santiago 19, Casilla 19001, Chile \label{ins9}
\and Max Planck Institute for Radio Astronomy, Auf dem Hügel 69, 53121 Bonn, Germany, \label{ins10}
\and Sterrewacht Leiden, Leiden University, Postbus 9513, NL-2300 RA Leiden, the Netherlands \label{ins11}
\and Departments of Physics and Astronomy, Le Conte Hall, University of California, Berkeley, CA 94720, USA \label{ins12}
\and CENTRA, Centro de Astrofísica e Gravitação, Instituto Superior Técnico, Avenida Rovisco Pais 1, P-1049 Lisboa, Portugal \label{ins13}
\and Department of Astrophisical \& Planetary Sciences, JILA, Duane Physics Bldg., 2000 Colorado Ave, University of Colorado, Boulder, CO 80309, USA \label{ins14}
\and Department of Particle Physics \& Astrophysics, Weizmann Institute of Science, Rehovot 76100, Israel \label{ins16}
\and Institute of Astronomy, Madingley Road, Cambridge CB3 0HA, UK \label{ins17}
\and Instituto de Astronomía, Universidad Nacional Autónoma de México, Apdo. Postal 70264, Ciudad de México 04510, Mexico \label{ins18}
}

\date{Submitted 23 September 2020 / Accepted 21 April 2021 }
\abstract
{\object{GCIRS~7}, the brightest star in the Galactic central parsec, formed $6\pm2$~Myr ago together with dozens of massive stars in a disk orbiting the central black-hole. It has been argued that GCIRS~7 is a pulsating body, on the basis of photometric variability.}
{Our goal is to confirm photospheric pulsations based on interferometric size measurements to better understand how the mass loss from these massive stars enriches the local interstellar medium.}
{We present the first medium-resolution ($R=500$), K-band spectro-interferometric observations of GCIRS~7, using the GRAVITY instrument with the four auxiliary telescopes of the ESO VLTI. We looked for variations using two epochs, namely 2017 and 2019.}
{We find GCIRS~7 to be moderately resolved with a uniform-disk photospheric diameter of $\theta^*_\text{UD}=1.55 \pm 0.03$~mas ($R^*_\text{UD}=1368 \pm 26$ $R_\sun$) in the K-band continuum. The narrow-band uniform-disk diameter increases above 2.3~$\mu$m, with a clear correlation with the CO band heads in the spectrum. This correlation is aptly modeled by a hot ($T_\text{L}=2368\pm37$~K), geometrically thin molecular shell with a diameter of $\theta_\text{L}=1.74\pm0.03$~mas, as measured in 2017. The shell diameter increased ($\theta_\text{L}=1.89\pm0.03$~mas), while its temperature decreased ($T_\text{L}=2140\pm42$~K) in 2019. In contrast, the photospheric diameter $\theta^*_\text{UD}$ and the extinction up to the photosphere of GCIRS~7 ($A_{\mathrm{K}_\mathrm{S}}=3.18 \pm 0.16$) have the same value within uncertainties at the two epochs.}
{In the context of previous interferometric and photo-spectrometric measurements, the GRAVITY data allow for an interpretation in terms of photospheric pulsations. The photospheric diameter measured in 2017 and 2019 is significantly larger than previously reported using the PIONIER instrument ($\theta_*=1.076 \pm 0.093$~mas in 2013 in the H band). The parameters of the photosphere and molecular shell of GCIRS~7 are comparable to those of other red supergiants that have previously been studied using interferometry. The extinction we measured here is lower than previous estimates in the direction of GCIRS~7 but typical for the central parsec region.}


\maketitle

\section{Introduction}
The stellar population of the central parsec of the Galaxy has been widely studied \citep[and references therein]{2010RvMP...82.3121G, 2012RAA....12..995M}, where the presence of a disk of young stars is well recognized \citep{2000MNRAS.317..348G, 2003ApJ...594..812G, 2006ApJ...643.1011P,  2009ApJ...690.1463L, 2009ApJ...697.1741B, 2014ApJ...783..131Y}. Most of these stars are massive O-type supergiants and Wolf-Rayet stars \citep{2007A&A...468..233M, 2010ApJ...708..834B, 2014A&A...567A..21S}. GCIRS~7 is one of the few evolved late-type stars \citep[an M1 red supergiant or RSG,][]{1996AJ....112.1988B} and a SiO maser source \citep{1997ApJ...475L.111M} as well as the brightest star (in the H and K bands, with  $H=9.5$ and $K=6.5$) of all the central parsec {\citep{1975ApJ...200L..71B}}. The works of \citet{1991ApJ...371L..59Y}, and \citet{1991ApJ...378..557S}  reported a cometary tail whose origin comes from GCIRS~7; more recently, \citet{2020PASJ..tmp..162T} revealed the presence of an ionised shell in the core of the cometary tail, estimating the mass loss of GCIRS~7 with ALMA observations.

This stellar population is permeated by the complex interstellar medium (ISM) environment and interacts with it. The central parsec is surrounded by a 2--7~pc-wide clumpy torus, the Circumnuclear Disk (CND), composed of dust and neutral gas \citep{1982ApJ...258..135B}. The \ion{H}{ii} region Sgr~A West \citep[the Minispiral; e.g.,][]{1980ApJ...241..132L, 1983Natur.306..647L} consists of tidally-sheared streamers and smaller patches and filaments of dust and ionised gas that orbit and penetrate the central parsec \citep{2003A&A...408.1009L, 2004A&A...426...81P, 2007A&A...469..993M, 2012ApJ...755...90I, 2017ApJ...842...94T}. The volume surrounding these components inside the central cavity of the CND is not empty but, rather, filled with hot ($\approx 1.3$~keV) plasma detected in X-ray \citep{2003ApJ...591..891B}. Finally, warm H$_2$ (with an excitation temperature of $T_\text{e}\approx 2000$~K) has been detected throughout the central parsec, presumably at the surface of many dusty clumps \citep{2016A&A...594A.113C, 2019A&A...621A..65C}. \citet{2012A&A...540A..50F} gives an interesting overview of the ISM content of the central parsec. The large and dense clumps that form the Minispiral are believed to be infalling from the CND and beyond, but the origin of the lighter and less dense features (filaments, smaller clumps, X-ray plasma) is less clear. A fraction of them could originate in the feedback from the massive stars. G1 \citep[also known as Sgr~A*-f,][]{2004A&A...417L..15C, 2005A&A...439L...9C} and G2 \citep{2012Natur.481...51G} may well be extreme examples of such feedback clumplets \citep[e.g.,][]{2012A&A...546L...2M, 2014ApJ...789L..33D, 2015ApJ...811..155S}.

Thanks to the performance of stellar interferometers, the understanding of the structure and evolution of RSG has improved significantly. The closest ones have been widely studied, not only obtaining measurements  of their sizes but also revealing single-layer atmospheres \citep{2005A&A...436..317P, 2007A&A...474..599P, 2014A&A...572A..17M}, multi-layer atmospheres \citep{2009A&A...503..183O, 2011A&A...529A.163O, 2013A&A...555A..24O, 2019MNRAS.489.2595H}, complex structures in the photosphere \citep{2009A&A...508..923H, 2010A&A...511A..51C, 2011ApJ...740...24R,2017A&A...602L..10O, 2017Natur.548..310O}, and even the temporal evolution of the stellar surface \citep{2011A&A...529A.163O, 2013A&A...555A..24O, 2016A&A...588A.130M, 2018A&A...614A..12M, 2020A&A...635A.160C}. Moreover, imaging of a RSG was performed in \citet{2014ApJ...785...46B}, \citet{2014SPIE.9146E..1QM}, and more recently, in \citet{2017A&A...606L...1W} and \citet{2020A&A...635A.160C}. Although the sample of spatially resolved RSGs has been increasing over the last decade \citep{2013A&A...554A..76A, 2015A&A...575A..50A, 2017A&A...597A...9W}, this sample is still not very large due to the shortness of the RSG phase, hence, only a small number of stars can be resolved using interferometers. When available, the study and characterization of the outer atmosphere of any new RSG and its temporal evolution would add valuable knowledge to the understanding of their mass loss processes, which have not yet been fully described from the first principles \citep{2020MNRAS.492.5994B}.

\citet{2014A&A...568A..85P} compiled almost 40 years of near-infrared photometric data on GCIRS~7 and exhibited two periods in the light curves: a short ``fundamental'' period, $P_0\approx470$ days, and a long ``secondary'' period, $P_\text{LSP}\approx2800$ days, as are often seen in RSGs. Those periods are believed to be a sign of pulsations, especially for  the $P_0$ of the fundamental or first overtone radial mode \citep[and references therein]{2012ApJ...754...35Y}. Such pulsations are expected to play a major role in the mass loss of RSGs. However, they have never been confirmed on the basis of direct size measurements. GCIRS~7 has been observed via interferometry on the VLTI using AMBER in the K band and PIONIER in the H band \citep{2008A&A...487..413P, 2014A&A...568A..85P}; however the AMBER data do not have sufficient spectral resolution and $(u, v)$-coverage to disentangle the stellar disk from the circumstellar environment so that only the PIONIER data provide a trustworthy uniform-disk diameter ($\theta_\text{UD}(2013)=1.076\pm0.093$~mas).

The GRAVITY instrument has tremendously increased the sensitivity of the VLTI \citep{2017A&A...602A..94G}, allowing us to observe GCIRS~7 at moderate spectral resolution ($R=500$) in single-field mode using the four $1.8$-m auxiliary telescopes (AT) at two epochs (2017 and 2019), with the goal of detecting variations in the photospheric diameter of the star and in its circumstellar environment. The data sets, the data reduction, and the calibration processes are described in Sect.~\ref{sec:data}. The methods and models used to measure the parameters of the star are described in Sect.~\ref{sect:model}. The results and their implications are discussed in Sect.~\ref{sec:discuss}.  Finally, our conclusions are presented in Sect.~\ref{sec:conc}.

\section{Data}\label{sec:data}

\begin{figure}
  \centering
    \includegraphics[width=\linewidth]{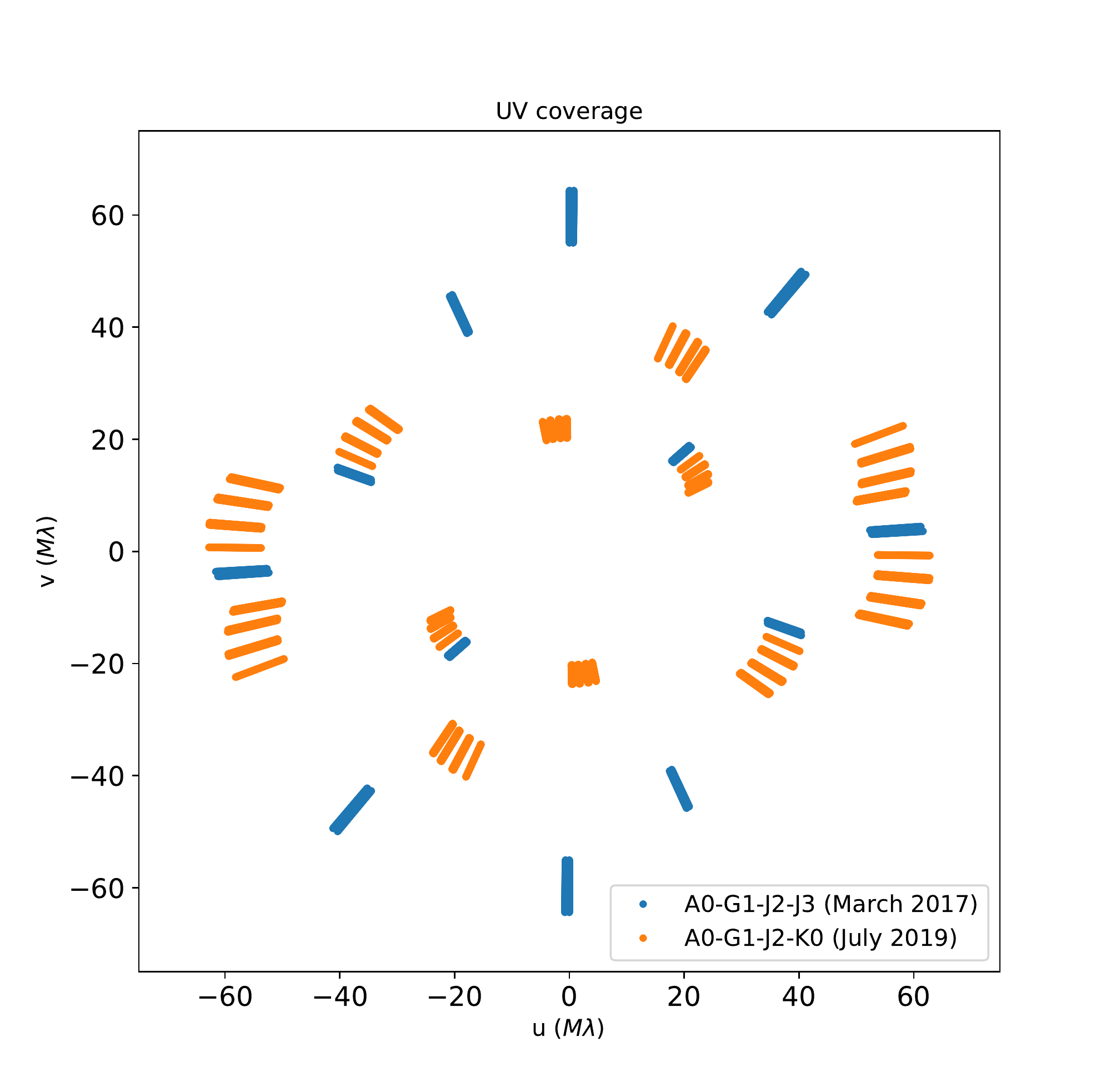} 
   \caption{(u,v) plane coverage for the two epochs 2017 and 2019. Each (u,v) point is elongated to account for the spectral range.} 
\label{fig:uvcoverage}
\end{figure}

The log of the data is presented in Table \ref{tab:log} with science on-target (SCI), calibrator on-target (CAL), and sky (SKY) frames. The data were taken at two different epochs (two SCI data frames with run ID 098.D-0250(B), corresponding to the night of 18 March 2017 and seven SCI data frames with run ID 103.B-0032(F), corresponding to the nights of the 5 and 6 July 2019) with two different baseline configurations, as shown in Figure \ref{fig:uvcoverage}. The maximum baseline was 132.5~m in 2017 and 129.3~m in 2019. Turbulence in the beams was corrected with the NAOMI adaptive optics (AO) system \citep{2016SPIE.9907E..20G, 2019A&A...629A..41W} on axis for the 2019 data, while only tip-tilt stabilization using STRAP on a nearby visible star was possible in 2017. Fringe-tracking was performed on-axis at both epochs. Only the science beam combiner data were used in this study since the fringe-tracker data do not have sufficient spectral resolution for our purpose.

We chose a single target to be the calibrator for both spectroscopy and interferometry. We used this calibrator to remove the atmospheric features using the appropriate template (see Sect.~\ref{sec:specal}). The calibrator is always observed at the same air mass and integration time as the science sequence ($\text{DIT}=10$~s, $\text{NDIT}=30$, $\text{air mass}=1.01$ in 2017 and $\text{DIT}=5$~s, $\text{NDIT}=30$, $\text{air mass}=1.01$ in 2019). We recorded the 2017 data in combined polarization mode and the 2019 data in split polarization mode, making use of the Wollaston prism. For this reason, the second run in 2019 results in two simultaneously-recorded data sets, one for each polarization (P1 and P2).

\begin{table}
\caption{Observation log. HD45124 was used for spectral calibration only. Atmospheric data were obtained from Paranal ASM (Astronomical Site Monitoring).} \label{tab:log}
\centering
\resizebox{\linewidth}{!}{ 
\begingroup
\setlength{\tabcolsep}{10pt}
\renewcommand{\arraystretch}{1.5}
\begin{tabular}{| c | c | c | c |}
\hline
Time (UT)    &  Target       & Seeing (")  & $\tau_0$ (ms) \\ \hline
\multicolumn{4}{|c|}{18-03-2017 (COMBINED mode)}\\
 \hline 
09:43:10   & GCIRS~7   (SCI-SCI-SKY)          & 0.55-0.58    & 6.27      \\ 
10:08:04   & HD160852 (CAL-SKY)  & 0.71-0.64    & 4.48      \\ 
 \hline
\multicolumn{4}{|c|}{05-07-2019 (SPLIT mode)}\\
 \hline 
04:49:38   & GCIRS~7 (SCI-SCI-SKY)    & 0.69-0.53    & 4.49      \\
05:08:29   & HD161703 (CAL-CAL-SKY)   & 0.72-0.81   & 4.45      \\
 \hline 
\multicolumn{4}{|c|}{06-07-2019 (SPLIT mode)}\\
 \hline 
03:13:00  & GCIRS~7  (SCI-SKY)   & 0.51-0.60         & 6.66      \\
03:25:51  & HD161703 (CAL-CAL-SKY)   & 0.60 -0.64     & 5.59      \\
03:43:00  & GCIRS~7    (SCI-SCI-SKY)   & 0.53-0.55    & 5.88      \\
03:59:51  & HD161703  (CAL-CAL-SKY)   & 0.72-0.88      & 3.68      \\
04:17:12  & GCIRS~7    (SCI-SCI-SKY)   & 0.91-0.75       & 3.55      \\
04:33:39  & HD161703  (CAL-CAL-SKY)   & 0.73-0.79      & 5.23     \\     \hline
\end{tabular}
\endgroup
}
\end{table} 

\subsection{Data reduction and calibration}

\subsubsection{Interferometric quantities}
The data were reduced with the GRAVITY pipeline \citep{2014SPIE.9146E..2DL}. As the source is moderately resolved, both the closure phase signal and the differential phase are $0\degr\pm3\degr$ and we only used the squared visibilities ($V^2$) for the interferometric analysis. The pipeline was also used to produce the photometric spectra. The squared visibilities were calibrated with HD160852 \citep[$\theta_\mathrm{LD}=0.148\pm0.004$ mas, $V^2=0.997\pm0.001$, ][]{2016A&A...589A.112C} in 2017 and HD161703 \citep[$\theta_\mathrm{LD}=0.38\pm0.01$ mas, $V^2=0.983\pm0.001$,][]{2016A&A...589A.112C} in 2019, where $\theta_\mathrm{LD}$ is the limb-darkened diameter, which, in the original work, was obtained via a polynomial fitting. Both calibrators are observed at the same air mass as GCIRS~7.

\subsubsection{Photometric spectra calibration}\label{sec:specal}

We used several templates from \citet{1998PASP..110..863P} to perform the absolute flux calibration of the spectra. In order to get a robust estimate of the error bars related to the template choice, we chose templates A0V, A3V, F2V, and F5V for 2017 and K0III, K2III, and K3III for 2019. The following formula yields the calibrated flux density of GCIRS~7, measured in $\mathrm{W cm^{-2}\mu m^{-1}}$:

\begin{equation}
F_{\lambda}^\text{IRS7}= \frac{F(\lambda)^\text{IRS7}_\text{data}}{F(\lambda)^\text{cal}_\text{data}} \times{F(\lambda)^\text{template}} \label{eq:Firs7}\text{ ,}
\end{equation}

\noindent where $F(\lambda)^\text{template}$ is the corresponding spectral template from \citet{1998PASP..110..863P} for each epoch (four templates in 2017, three templates in 2019). For each template, a spectra $F_{\lambda}^\text{IRS7}$ is obtained. We take the average of these individual estimates as our final calibrated spectrum for each epoch, and their standard deviation as the uncertainty on this calibration.

While the 2017 data were recorded in combined polarization mode resulting in a single spectrum, the 2019 data were recorded using the Wollaston prism, resulting in two independent spectra which were calibrated individually and then co-added into a single spectrum. 

\section{Modeling}\label{sect:model}

\subsection{Local interstellar extinction and spectrum normalization}\label{sec:locext}

It is possible to measure the effect of the interstellar extinction in the direction of the source using the slope of the continuum emission and considering an extinction law. Here, we consider the law derived in \citet{2011ApJ...737...73F}:

\begin{equation}\label{eq:extinction}
    A_{\lambda}=A_0\times \left( \frac{\lambda}{\lambda_0} \right)^{\alpha}
,\end{equation}

\noindent with $\alpha=-2.11 \pm 0.06, \lambda_0=2.166$ $\mu \mathrm{m}$ (Brackett $\gamma$). An average value of $A_0=2.62\pm 0.11$ is given in \citet{2011ApJ...737...73F} for the Galactic Center. However, it is known 
that extinction varies significantly throughout the central parsec, from a line-of-sight to the next and along a given line-of-sight \citep[e.g.,][]{2019A&A...621A..65C}. We therefore decided to derive the extinction from the source itself. Starting from a gray atmosphere approximation ($F_{\lambda}\propto B(\lambda, T) 10^{-0.4 A_{\lambda}}$), an expression for $A_0$ can be obtained by introducing $A_\lambda$ from Eq.~\ref{eq:extinction} and taking the derivative of the logarithm of $F_\lambda$ with respect to $\lambda$:

\begin{equation} \label{eq:AO}
   A_0=\frac{-2.5}{\alpha \log{(10)}}\left(\frac{\lambda}{\lambda_0}\right)^{-\alpha}\left[\lambda\frac{\partial F_{\lambda}}{\partial \lambda}- \frac{hc}{\lambda k T}\left(1+\frac{1}{e^{(hc/\lambda k T)}-1}\right)-5\right]
,\end{equation}\noindent where $h,k,c$ are the Planck and Boltzmann constants and the speed of light respectively. We used the effective temperature determined by \citet{2014A&A...568A..85P} ($T=3600  \pm 195 $~K). The derivative is obtained for each spectrum in the sample (8 in 2017, 56 in 2019) through a linear regression in the continuum sub-band ($\lambda=2.1-2.2$~$\mu\mathrm{m}$), yielding as many estimates of $A_0$. The average of those individual estimates gives our final estimate for each epoch:

\begin{itemize}
    \item 2017: $A_0=3.26 \pm 0.35;$
    \item 2019: $A_0=3.28 \pm 0.26.$
\end{itemize}

To estimate the uncertainties, in addition to the flux error bars, we considered the deviation of the SED of a typical RSG from a blackbody. For this purpose, we used the de-reddened and absolute-calibrated spectrum of Betelgeuse from \cite{2009ApJS..185..289R} to estimate the deviation. This star shares the spectral class with GCIRS~7 and was found to have a same effective temperature, which has not substantially changed in recent decades (several measurements are presented in Sec.~\ref{sec:Molspheres}). An estimate of the slope for the sub-band used in our work to estimate the extinction ($\lambda=2.1-2.2$~$\mu\mathrm{m}$) gives a deviation of 11\% while comparing the slope of the blackbody with a linear fit of the spectrum of Betelgeuse in the continuum sub-band. This deviation in the slope is propagated by using Eq.~\ref{eq:AO}, giving an error bar of 0.20. This value has been considered in addition to the error propagation of the fluxes to estimate the uncertainty.

The two values are compatible and can be averaged down to a single estimate of the extinction: $\bar{A}_{0}=3.27 \pm 0.20$. We can use this value to de-redden $F^\text{IRS7}_\lambda$ (Eq.~\ref{eq:Firs7}) and integrate over the $K_S$ band to determine the broad-band extinction specifically for the photosphere of GCIRS~7:

\begin{equation}A_{K_S} = -2.5\log{\frac{\int_{K_\text{S}} F_\lambda^\text{IRS7} d\lambda}{\int_{K_\text{S}} F_\lambda^\text{IRS7} \times 10^{+0.4A_\lambda}d\lambda}}=3.18\pm0.20.
\end{equation}

Then each spectrum is normalized separately using the average $A_0$ for each epoch. Finally, a weighted average of the normalized spectra is computed for each epoch, the weights being the inverse of the square of the error bars. The weighted average of the normalized spectra is noted as $F_{\lambda}^\text{N}$ throughout this work.

\subsection{Squared visibilities}
\label{sec:UDa}
\begin{figure}
\centering
\includegraphics[width=\linewidth]{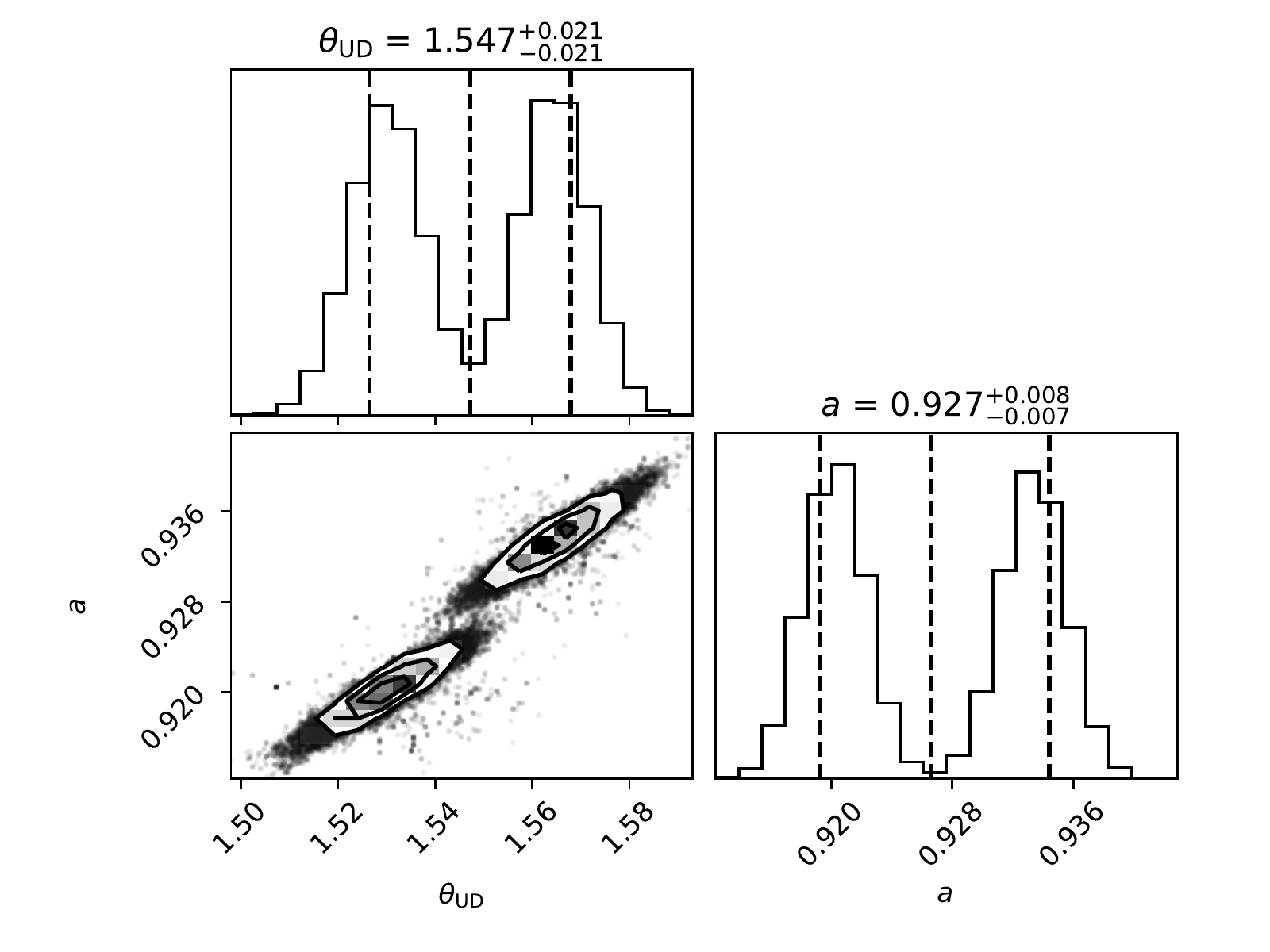}
\includegraphics[width=\linewidth]{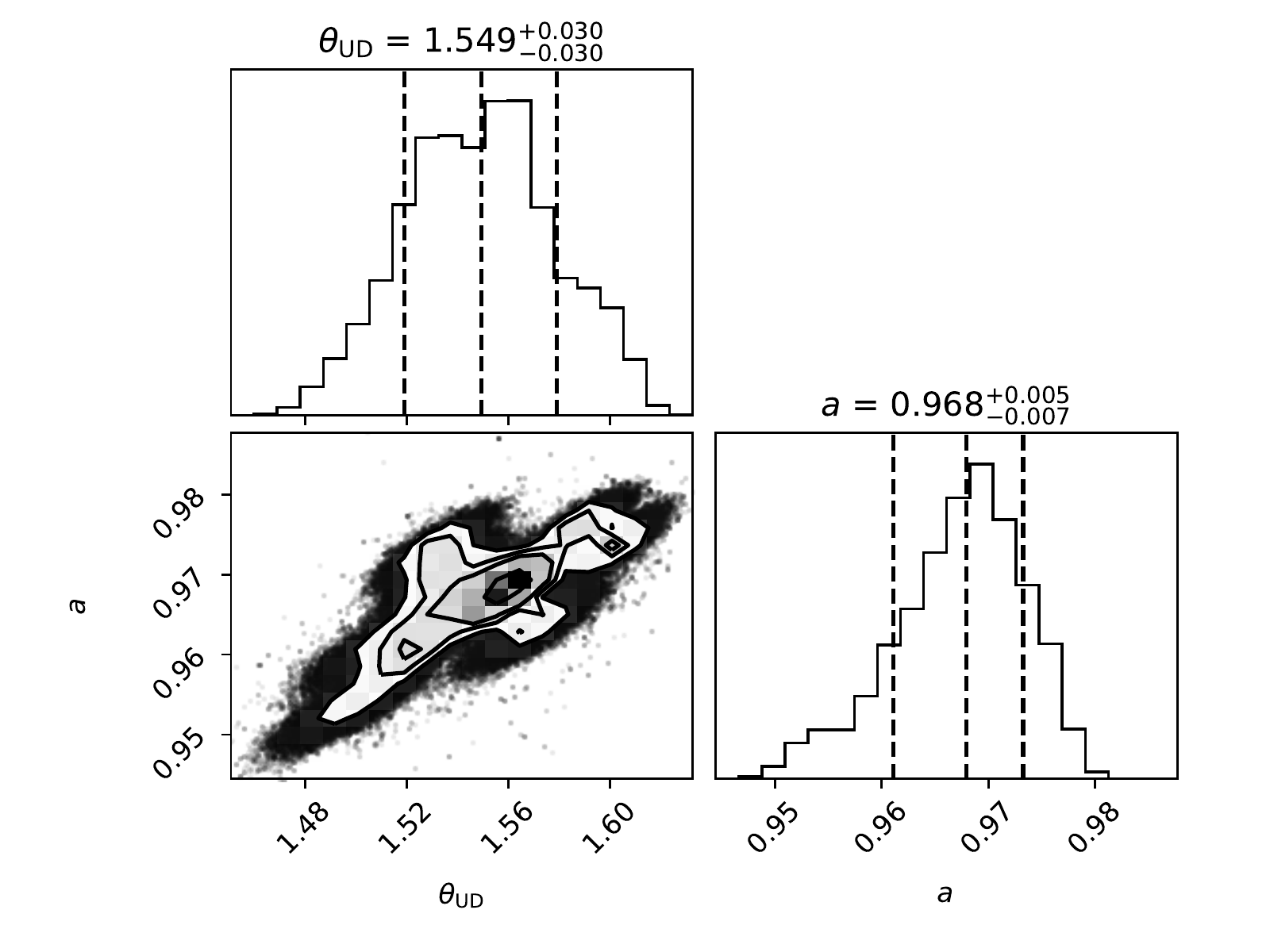}
\caption{Combined corner plot of the uniform-disk plus background MCMC samples on each data file for 2017 (top) and 2019 (bottom).}\label{fig:cornerplots}
\end{figure}

\begin{figure}
\centering  
\includegraphics[width= \linewidth]{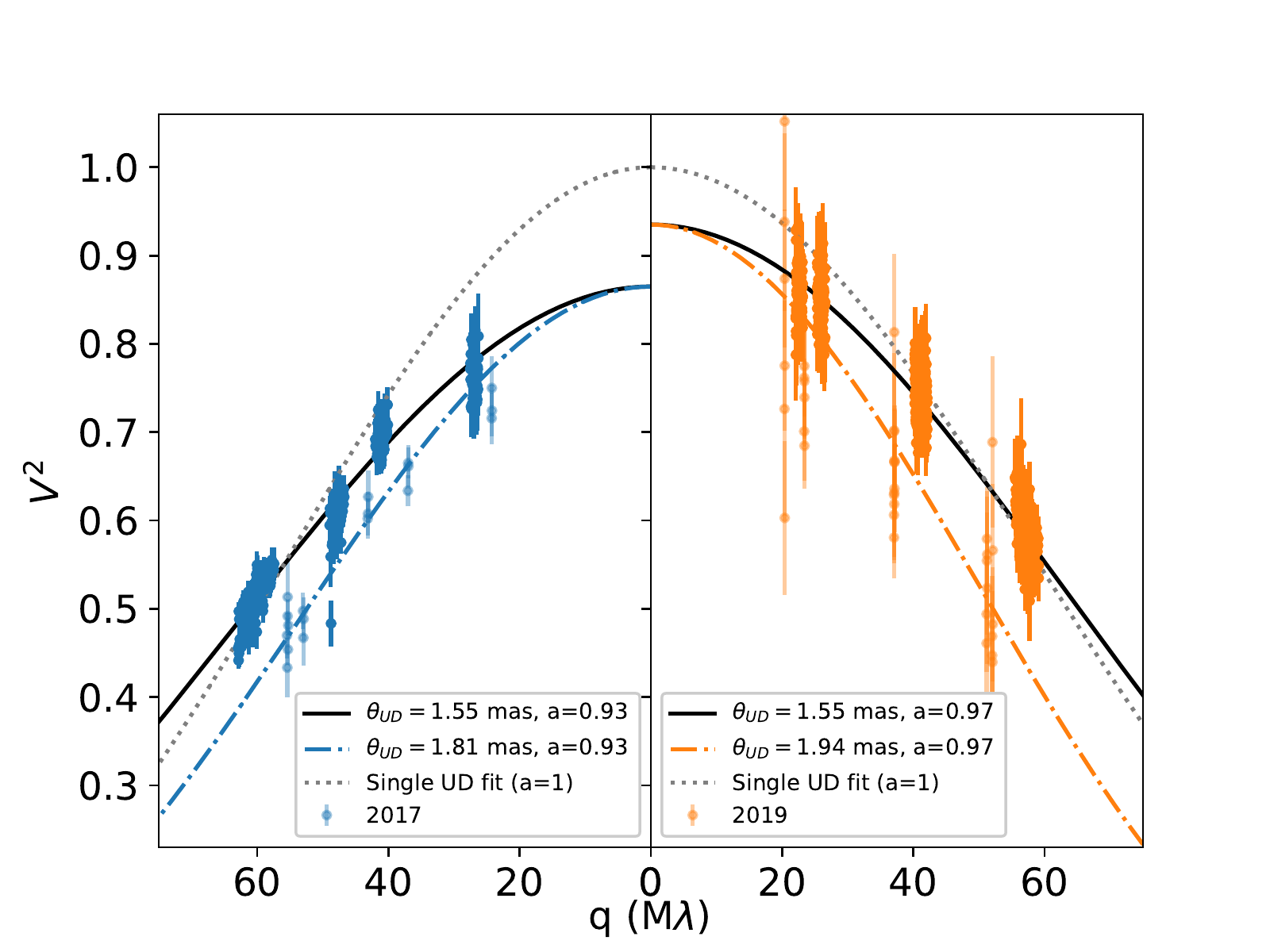}
\caption{$V^2$ measurements of 2017 and 2019. Two uniform-disk plus background models are shown: the average best-fit in the continuum (2.1--2.2$\,\mu$m, black line) and the average uniform-disk diameter in the CO 5-3 band head ($2.381$--$2.387\,\mu$m, colored lines), both of them with the measured background level in parenthesis. Only the data in those two wavelength ranges are shown. The dotted line reveals the pure UD fit of the visibilities in the continuum assuming no incoherent background.}\label{fig:viscontpeak} 
\end{figure}

\begin{figure*}
\centering  
\includegraphics[width=\linewidth]{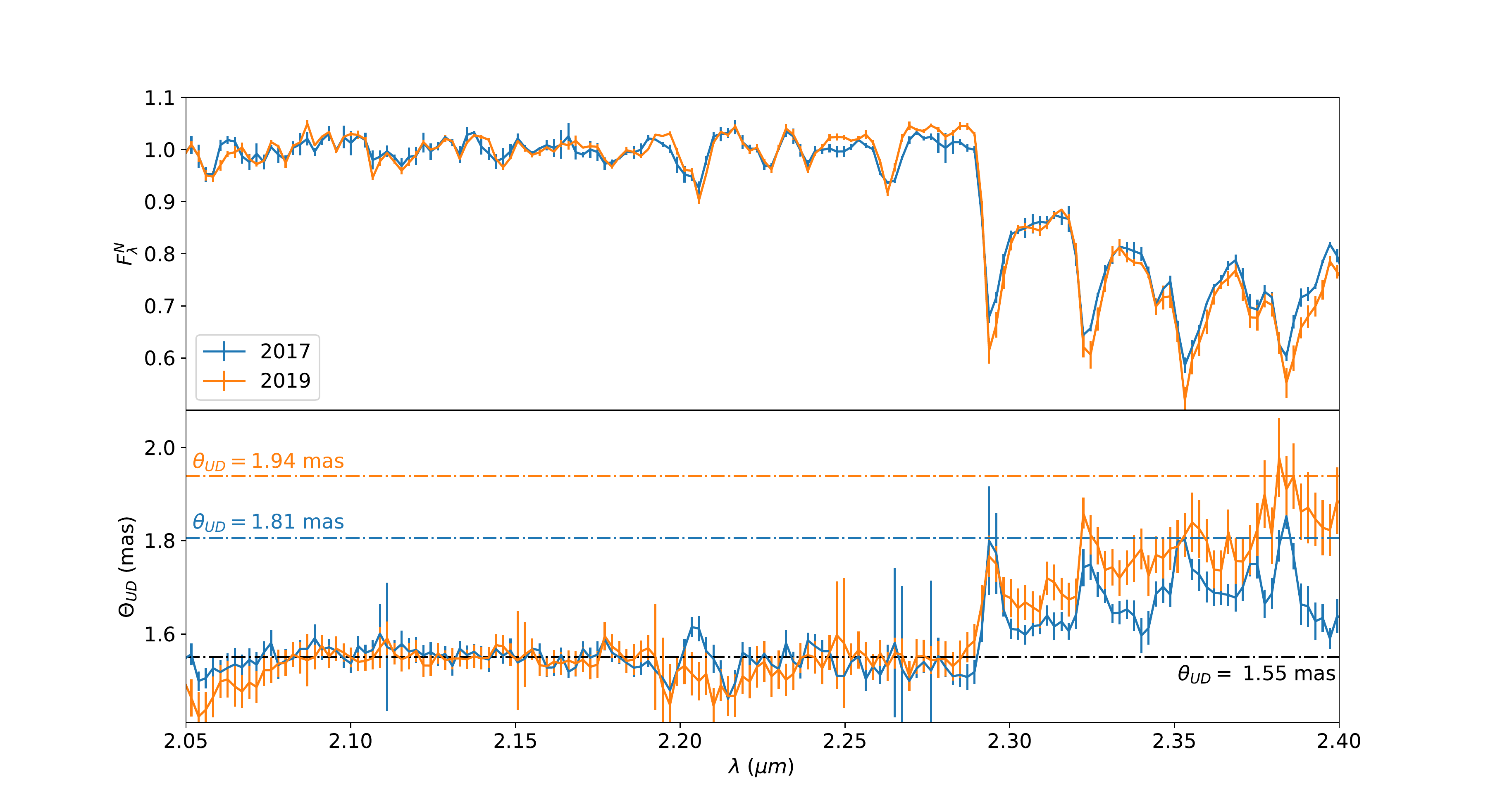}
\caption{Normalized spectra  and uniform disk diameters for the two epochs. In 2019, the uniform disk profile is the average of the profiles of the two polarization states.}\label{fig:FnUD1719}
\end{figure*}

The target is only moderately resolved by GRAVITY ($V^2>0.4$) and therefore the data are not sensitive to limb darkening. We modeled the visibility in the continuum ($2.1$--$2.2\,\mu$m) with a uniform disk plus a spatially over-resolved background as in \citet{2007A&A...474..599P}:

\begin{equation}\label{eq:UDfit} 
V_\text{aD}^2(q, a, \theta^*_\text{UD})=a^2 V_\text{UD}^2(q, \theta^*_\text{UD})
,\end{equation}

where $V_\text{aD}(q, a, \theta^*_\text{UD})$ is the visibility of this uniform-disk plus the spatially over-resolved background, $V_\text{UD}(q, \theta^*_\text{UD})$ is the usual uniform-disk visibility, $q=\sqrt{u^2+v^2}$ is the spatial frequency ($u$ and $v$ are the spatial frequency components), $\theta^*_\text{UD}$ is the angular diameter of the disk, and $a$ is the fraction of stellar flux over the total injected flux. 

To explore the wavelength dependence of the parameter $a$, we performed a uniform-disk plus background fit for each spectral channel in the continuum region (2.1-2.2 $\mu$m). The results for $a$ do not reveal any significant wavelength dependency and therefore we assumed the stellar flux and background to exhibit the same spectrum. 

We estimated $\theta^*_\text{UD}$ and $a$ using a Monte-Carlo Markov chain (MCMC) algorithm based on the Python package \texttt{emcee} \citep{2013PASP..125..306F}. We let 100 walkers evolve for 600 steps for each file and each polarization state (for 2019) individually, using all wavelengths from 2.1 to 2.2~$\mu$m. The individual Monte-Carlo simulations (2 in 2017, 14 in 2019) yield distributions that are too far apart compared to their internal scatter (Fig.~\ref{fig:cornerplots}). This is because systematic errors resulting from the variation in the transfer function dominate over statistical errors. For this reason, we combined all the samples from these simulations in a single histogram per epoch. The 2017 combined histogram is clearly bi-modal because there are only two individual frames, while the 2019 data set is rich enough that the 14 individual Gaussian-like histograms merge into a broad Gaussian-like peak. The two polarization states in the 2019 data give similar distributions and are combined together. As our best estimate for the two parameters, we use the median of the 1D combined histograms (which is equivalent to taking the average of the median values of the individual 1D histograms). The 2019 data set is well suited to determine the uncertainties because it contains enough individual measurements for their scatter to make sense. Using the $16\%$ and $84\%$ percentiles of the 1D histogram yields uncertainties that combine statistical errors, instrumental stability, and degeneracy between the two parameters. The same approach on the 2017 data leads to a value for the uncertainty on $\theta^{*}_\text{UD}$ that we deem too small, being dominated by the scatter of two individual values. We therefore adopt the 2019 error bar instead also for 2017. Our final estimates for the two years are as follows:
\begin{itemize}
\item 2017: $a=0.927 \pm 0.008$, $\theta^{*}_\text{UD}=1.547\pm 0.030$~mas;
  \vspace{1mm}
\item 2019: $a=0.968 \pm 0.006$, $\theta^{*}_\text{UD}=1.549\pm 0.030$~mas.
\end{itemize}

It is worth noting that this uncertainty on the uniform disk diameter ($30\;\mu$as) is itself quite an achievement, on the same order as the astrometric measurements performed by GRAVITY \citep[e.g.,][]{2020A&A...636L...5G}. These average best-fit models are shown in Fig.~\ref{fig:viscontpeak}., compared with the best single UD fit without considering the background.

The fraction of coherent flux $a$ means that the fibers of GRAVITY were fed with $7\%$ incoherent light from the circumstellar background in 2017 and $3\%$ in 2019. This decrease may be due to the improved turbulence correction offered by the AO system NAOMI compared to tip-tilt stabilization with STRAP, since the residuals of this correction directly translate into a widening of the fibre field-of-view \citep{2019A&A...625A..48P}. We removed this background from the visibilities by dividing them by $a$ for each epoch. The use of a single value instead of a value per spectral channel is justified as the spectral resolution is moderate and the flux ratio between the emission of the circumstellar dust and stellar flux varies only slowly across the K band. 

After correcting for this background, a simple uniform-disk fit was performed for each spectral channel in the interval ($2.05$--$2.4\, \mu$m). The results are displayed in Fig.~\ref{fig:FnUD1719}, showing both the normalized spectra derived in Sect.~\ref{sec:locext} and the chromatic uniform-disk diameter profile as a function of wavelength. The diameter is close to constant in the continuum between 2.1 and 2.2$\,\mu$m, again showing that the continuum is very well fitted by a uniform-disk plus the spatially over-resolved background. The unbiased weighted standard deviation of $\theta_\text{UD}$ through the continuum (again 2.1--2.2~$\mu$m) is 0.020~mas in 2017 and 0.018 in 2019. At this level, the main source of uncertainty on $\theta^*_\text{UD}$ is the presence of many absorption lines in the spectrum, which form at various altitudes in the atmosphere of the star. Therefore, the limitation is fundamental: it is the definition of the photosphere itself.

The average value of $\theta_\text{UD}$ above 2.29~$\mu$m is 1.66~mas in 2017 and 1.77~mas in 2019, that is, a 12-$\sigma$ departure from the continuum level. This sharp increase follows the absorption features that can be seen in the spectrum, with local maxima matching the deep CO band heads. Such variations correlated with the CO optical depth point towards a molecular shell above the photosphere as previously evidenced, for instance, by \citet{2004A&A...426..279P}, \citet{2005A&A...436..317P}, or \citet{2019MNRAS.489.2595H} for other evolved stars. 

In addition to the maxima in the CO bands, the uniform disk diameter increases monotonically after 2.29~$\mu$m in 2019. Such a feature has also been observed in Betelgeuse \citep{2014A&A...572A..17M} and other RSGs \citep{2013A&A...554A..76A, 2015A&A...575A..50A} and it can be attributed to the presence of H2O \citep{2014A&A...572A..17M}. However, this monotonic increase in the angular size is not clear in 2017, which hints at a variation in the stellar atmosphere between the two epochs.

A lower limit on the diameter of the shell is given by the maximum uniform-disk size in the molecular band: $\theta_\text{S} \geq \theta^\text{CO}_\text{UD}$, which we estimated by taking the average of the three $\theta_{UD}$ values around the CO 5-3 band head on each epoch: 
\begin{itemize}
\item 2017: $\theta^\text{CO}_\text{UD}=1.805 \pm 0.017\,$mas $=(1.166 \pm 0.027) \,\theta^{*}_\text{UD}$
  \vspace{1mm},
\item 2019: $\theta^\text{CO}_\text{UD}=1.939 \pm 0.019\,$mas $=(1.251 \pm 0.026) \,\theta^{*}_\text{UD}$,\end{itemize}
again neglecting correlated sources of error for this relative measurement.

This result provides evidence for the need to carry out a more complex model than that with a grey atmosphere to describe the physics of the object. In the next section, we model the molecular features with a simple shell model, which will also allow us to give a proper interpretation to the time variability which we see.

\begin{figure}
        \centering  
                \includegraphics[width=0.7\linewidth]{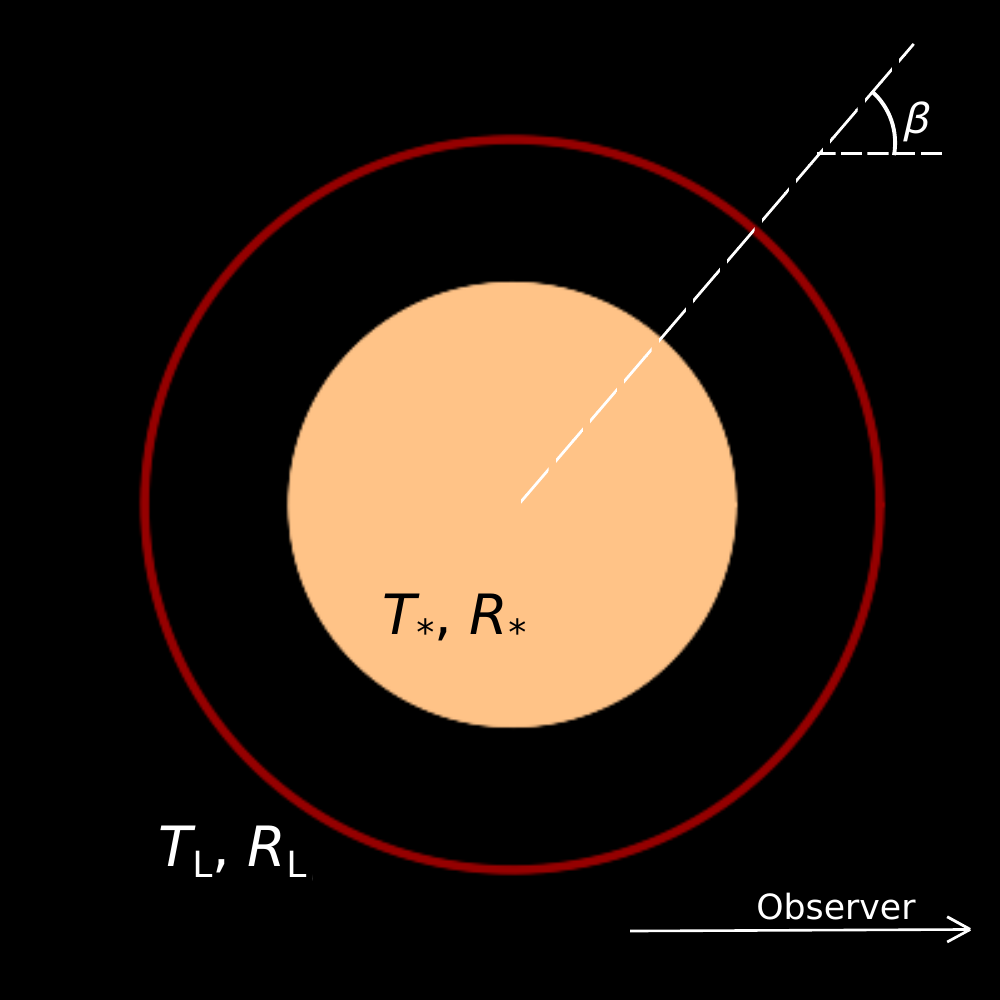}
        \caption{Sketch of the single-layer shell model.} 
\label{fig:MOLmodel}
\end{figure}

\subsubsection{Single-layer shell model}\label{sec:MOLsphere}

A geometrically thin molecular layer model has proven successful to reproduce the visibilities of other RSG stars \citep{2004A&A...426..279P, 2005A&A...436..317P, 2014A&A...572A..17M} and this is what we chose to use for this study. It has allowed for the interpretation of interferometric observations of RSG stars surrounded by a so-called MOLsphere according to the term coined by \citet{Tsuji2000a}. In this model, the shell is modeled with a temperature and an optical thickness but has zero geometrical thickness.

Both the stellar photosphere and the shell are modeled with black body functions. The specific intensity at angular distance $r$ from the center of the star as seen from the observer at wavelength $\lambda$ is given by:

\begin{equation}
 I^{r}_{\lambda}(T_*,T_{\text{L}},R_*, R_{\text{L}}, \tau_\lambda) =
   \begin{cases}
     B_\lambda(T_*)e^{(-\tau_\lambda/\cos{\beta})} \\
     + B_\lambda(T_\text{L})[ 1-e^{(-\tau_\lambda/\cos{\beta})}] & \text{if } r \leq R_* \\ \\
     B_\lambda(T_\text{L})[ 1-e^{(-2\tau_\lambda/\cos{\beta})}]  & \text{if } R_* < r \leq R_\text{L} \\ \\
     0 & \text{otherwise,} 
   \end{cases}\label{eq:Ilayermodel}
,\end{equation}

\begin{figure}
  \centering
    \includegraphics[width=\linewidth]{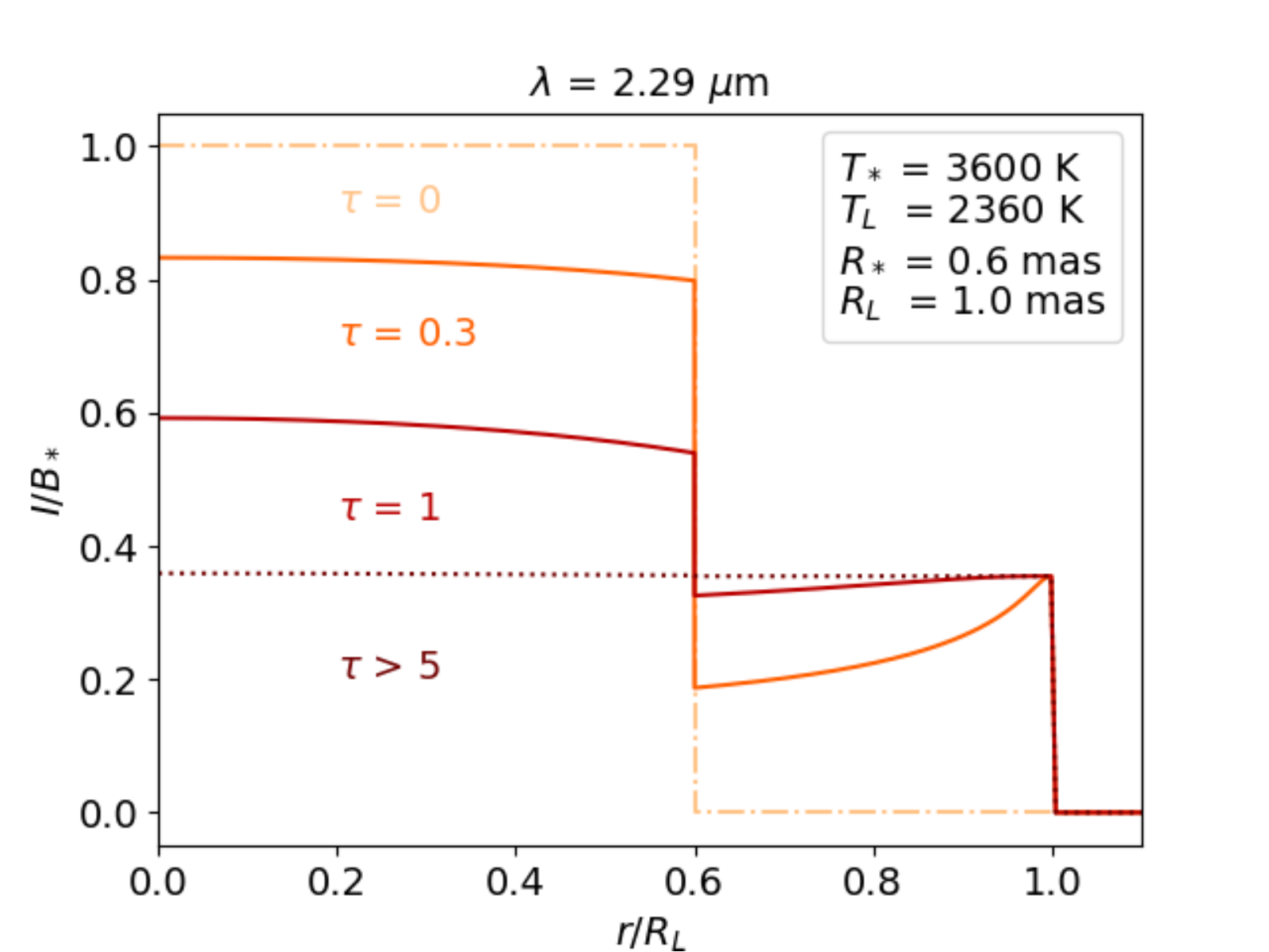} 
  \caption{Spatial intensity profile of the single layer shell model (Eq.~\ref{eq:Ilayermodel}) normalized by the photosphere black body function for various values of the optical depth of the shell $\tau$. The abscissa is relative to the size of the layer. Here,  the maximum attenuation reaches a factor 1.55 for a temperature of the layer $T_\text{L}=2360\,$K. }\label{fig:normI}
\end{figure}

\noindent where $T_*$ and $T_\text{L}$ are the temperatures of the photosphere and of the molecular layer, $R_*$ and $R_\text{L}$ are their angular radii (hence $R_*=\theta^*_\text{UD}/2$), $\tau_\lambda$ is the optical depth of the molecular layer at wavelength $\lambda$, $B_\lambda(T)$ is the Planck function at wavelength $\lambda$ and temperature $T$, and $\beta$ is the angle between the radius vector and the line-of-sight so that $\cos{\beta}=\sqrt{1-(r/R_\text{L})^2}$. A sketch of the model is displayed in Fig.~\ref{fig:MOLmodel}.

The center-to-limb variation is illustrated in Fig. \ref{fig:normI} for various optical depths. The sharp variation near $r/R_\text{L}=0.6$ corresponds to the edge of the photosphere, which is assumed to be a uniform disk. The increase of the projected layer optical depth with increasing $r$ (due to the $\cos\beta$ factor) causes a slight limb darkening on the disk over the photosphere,  and a limb brightening between the limb of the photosphere and the limb of the shell. The model fails to reproduce infinitely strong flux attenuation by the shell unless its temperature reaches 0\,K. For example, the maximum attenuation reaches a factor $1.55$ for a temperature of $T_\text{L}=2360\,$K and is typically reached for optical depths of 5 or larger. The fact that at high optical depth the shell itself behaves as a photosphere is one of the shortcomings of this simple model for the interpretation of the surface brightness in the core of molecular lines.

\begin{figure*}
  \centering
    \includegraphics[width=\linewidth]{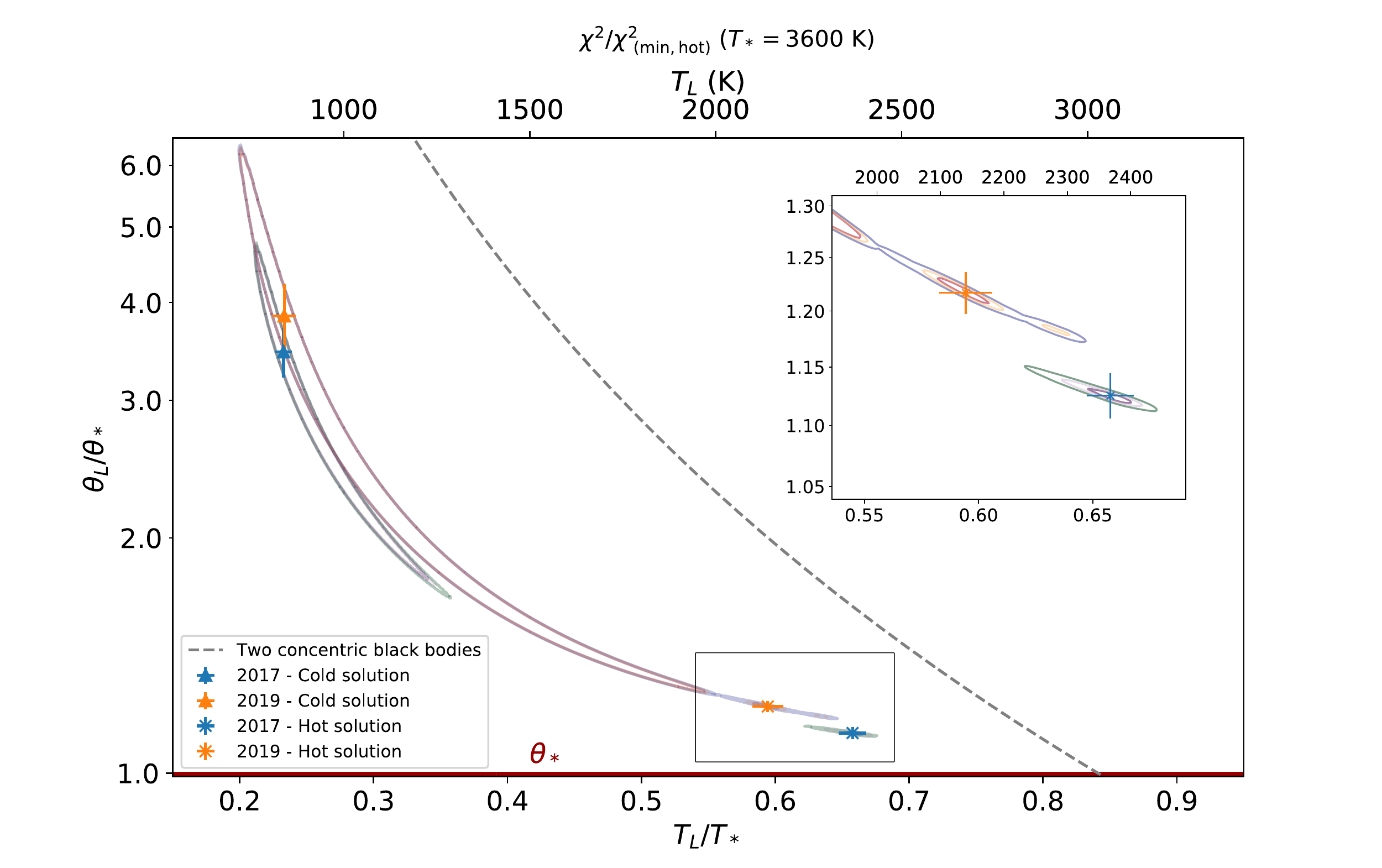}
  \caption{Contour plot of the $\chi^2$ maps (Eq.~\ref{eq:chiMOL}) considering $T_\text{L}, \theta_\text{L}$, which are the two parameters of interest. \emph{Solid lines:} Decrease of likelihood of 1, 3, and 5 $\sigma$ relative to the hot solution minimum of each epoch \citep{PresTeukVettFlan92}. \emph{Grey dashed line:} Temperature profile of a spherically thin shell enclosing a perfect black body (Eq.~\ref{eq:BBeq}). \emph{Triangles with error bars:}  Cold solutions for both epochs corresponding to the absolute minima. \emph{Crosses with error bars:}  Hot solutions for both epochs (corresponding to the local minima). }\label{fig:chimap}
\end{figure*}

We adopted the effective temperature of \citet{2014A&A...568A..85P} for the photosphere: $T_*=3600\,$ K. We also fixed the photospheric diameter to the continuum uniform-disk diameter from Sect.~\ref{sec:UDa}: $R_*=\theta^*_\text{UD}/2=0.775$~mas. In this section, we focus on the molecular features above 2.28 microns \citep{2014A&A...572A..17M}, which includes CO bands but also water vapor to obtain measurements for $T_\text{L}$ and $R_\text{L}$. 

A physical modeling of the wavelength-dependent optical depth of the thin layer $\tau_\lambda$, while possible, would significantly add to the complexity of the model without begin necessarily accurate because of the very simplistic geometrical model. However, for a set of parameters $(T_*, T_\text{L}, R_*, R_\text{L})$ and for each wavelength, the relation between $F_\lambda^N$ and $\tau_\lambda$ is bijective, except
where the model saturates as explained above. For any quadruplet $(T_*, T_\text{L}, R_*, R_\text{L})$ and for
each wavelength, we determine $\tau_\lambda$ univocally by finding the root of the following quantity:

\begin{equation}\label{eq:taus}
\left|F_{\lambda}^\text{N}-\frac{\int_{r=0}^{R_\text{L}}{I^{r}_{\lambda}(T_*,T_\text{L},R_*, R_\text{L}, \tau_\lambda)\times2\pi r dr}}{\int_{r=0}^{R_\text{L}}{I^{r}_{\lambda}(T_*,T_\text{L},R_*, R_\text{L}, \tau_\lambda=0) \times2\pi r dr}}\right|
,\end{equation}

which is implemented by a minimization since Eq.~\ref{eq:taus} is always positive. When the model does not saturate, this minimum reaches zero. When it does saturate (this happens in a few spectral channels of the CO band heads), the exact (large) value of $\tau_\lambda$ does not matter as the shell is then optically thick. A visibility model of the star and the shell $V^{2}_{\text{shell}}$ can then be built and compared to the corrected squared visibility data $V^2_{\text{c}, i}=V^2_i/a^2$, in which the effect of the incoherent background has been removed. This yields the following $\chi^2$:

\begin{equation}\label{eq:chiMOL}
\chi^2=\sum_{i=1}^{M}{\left(\frac{V^2_{\text{c}, i}-V^2_{\text{shell}}(q_i, T_*=T_\text{eff}, T_\text{L}, R_*=\theta^*_\text{UD}/2, R_\text{L}, \tau_{\lambda})}{\sigma_i}\right )^2}
,\end{equation}

where $M$ is the total number of data points considered for each epoch: $M=6 \times D \times n_\lambda$ where $D$ is the number of observations, $6$ the number of baselines and $n_\lambda$ the number of spectral channels in the band of interest. $T_\text{L}$ and $R_\text{L}$ are the only two remaining free parameters of the model. The fitting process involves then two steps:

\begin{itemize}
    \item For each pair $(T_L, R_L)$, $\tau_\lambda$ is measured from the spectrum for each wavelength by minimising Eq.~\ref{eq:taus}.
    \item By using the resulting set $(T_L, R_L, \tau_\lambda)$, the visibility squared of the model is computed and compared with the data via Eq.~\ref{eq:chiMOL}.
\end{itemize}

Figure~\ref{fig:chimap} displays a contour plot of this $\chi^2$ for each epoch in the $(T_\text{L}/T_*, \theta_\text{L}/\theta_*=R_\text{L}/R_*)$ plane (where $\theta_i=2R_i$). The map does not show a single clearly defined minimum, but a trough instead, thereby displaying a degeneracy between the radius and the temperature of the single layer of the shell model. For both 2017 and 2019, the trough has one global minimum at $(T_\text{L}, \theta_\text{L})=(838\pm 23\,\mathrm{K},5.36\pm 0.39\,\mathrm{mas})$ with $\chi^2_r=3.78$ in 2017 and $(T_\text{L}, \theta_\text{L})=(839 \pm 31\,\mathrm{K}, 5.96 \pm 0.58\,\mathrm{mas})$ with $\chi^2_r=1.98$ in 2019, which we refer to as the 'cold' solution and one local minimum at $(T_\text{L}, \theta_\text{L})=(2368 \pm 37\,\mathrm{K},1.74\pm0.03\,\mathrm{mas})$ with $\chi^2_r=3.93$ in 2017; and $(T_\text{L}, \theta_\text{L})=(2140\pm 42\,\mathrm{K},1.89\pm0.03\,\mathrm{mas})$ with $\chi^2_r=2.18$ in 2019, which we refer to as the 'hot' solution. In 2017, the two solutions are incompatible at 5$\sigma$.

Despite the fact that the cold solution corresponds to the absolute minimum, it cannot be the dominating solution of the model as its temperature is below the condensation temperature of silicate dust. In the context of mass loss, CO could be detected beyond the dust condensation radius as the CO gas can be dragged by the dust set in motion by the radiation pressure of the star. However, CO is naturally a major component of the molecular shell and therefore there should be an inner region where CO must be formed before being dragged by the dust. This inner shell must present a higher column density and a stronger spectroscopic signal in the K band than the cold solution. These requirements are satisfied by the hot solution.

In addition, we plotted in Fig.~\ref{fig:chimap} the position in radius and temperature for a fully absorbing and geometrically thin shell in thermal equilibrium above the photosphere:

\begin{equation}\label{eq:BBeq}
\frac{\theta_\text{L}}{\theta_*}=\frac{1}{\sqrt{2}}\left(\frac{T_*}{T_\text{L}}\right)^2
.\end{equation}

This model is more consistent with the hot solution than with the cold solution. For all these reasons, we believe the hot solution makes more physical sense than the cold solution although both are compatible with our data. However, the reduced $\chi^2$ of the hot solution is not ideal, which we interpret as the signature that our model is still too simple to perfectly reproduce the data. In an attempt to include these modeling errors into the statistical uncertainties, we rescaled the errors by the square root of the $\chi^2$ of their respective hot solution.

\begin{figure}
  \centering
    \includegraphics[width=\linewidth]{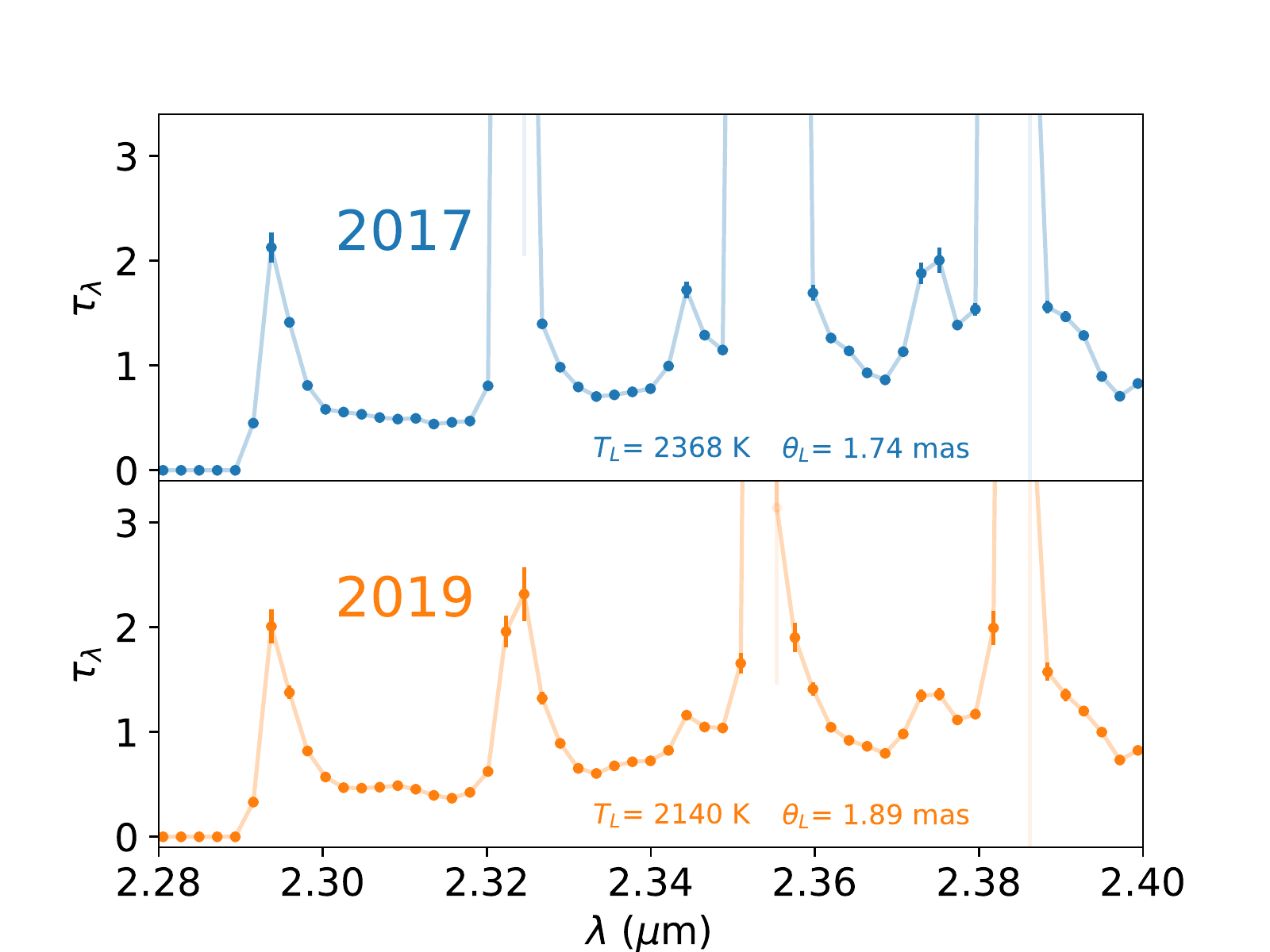} 
  \caption{Optical depths computed using Eq.~\ref{eq:taus} with the parameters of the hot solution. The peaks correspond to the CO band heads  where the single-layer model saturates as described in Sect.~\ref{sec:MOLsphere}.}\label{fig:tausch}
\end{figure}

The optical depths obtained for the hot solution of each epoch are presented Fig.~\ref{fig:tausch}. The error bars were obtained through Monte-Carlo error propagation in the $1\sigma$ contour of the hot solution. Only the values with $\tau\pm\Delta\tau<5$ are presented as the model reaches saturation near the peaks. A summary of the parameters of the model is presented in Table \ref{tab:results} for the hot solution. 

\begin{table}

\caption{Parameters of the single-layer molecular shell model for the hot solution.} \label{tab:results}
\centering
\begin{tabular}{| c | c | c |}
\hline
Parameter    & 2017       & 2019\\ \hline
$A_0$   & 3.27 (fixed)   & 3.27 (fixed)           \\ 
$\theta_*$ (mas) &  1.55 (fixed) & 1.55 (fixed)\\
$T_*$ (K) & 3600 (fixed) & 3600 (fixed)\\
$\theta_\text{L}$ (mas)  & $1.74 \pm 0.03$  & $1.89 \pm 0.03$          \\ 
$T_\text{L}$ (K)  & $2368\pm 37$     & $2140 \pm 42 $      \\ 
\hline
\end{tabular}
\end{table} 

A visibility contrast function is introduced to better show the effect of the molecular features on the visibilities. It is the average of the band to continuum visibility ratio over a subset of baselines: 

\begin{equation}\label{eq:Vcon}
C_{V2}(\lambda)=\left\langle\frac{V^2_c(q_i, T_*=T_\text{eff}, T_\text{L}, R_*=\theta^*_\text{UD}/2, R_\text{L}, \tau_{\lambda})}{V^2_c(q_i, T_*=T_\text{eff}, T_\text{L}, R_*=\theta^*_\text{UD}/2, R_\text{L}, \tau_{\lambda}=0)}\right\rangle_i
.\end{equation}\noindent

This contrast function, where the spectral and spatial information are combined in a synthetic way, is an interferometric counterpart to the normalized spectrum.

Figure~\ref{fig:Vcon} shows this function for the data of the two epochs as well as for synthetic data corresponding to the hot solution. For this plot, we chose the three partially overlapping longest baselines for each epoch. The visibility contrasts are plotted between  $2.28$ and $2.4\,\mu$m to show the upper part of the continuum and the molecular features. The single-layer shell model matches the data quite well except at the bottom of the band heads, where flux attenuation saturates with this simple model, as explained above.

\section{Discussion}\label{sec:discuss}
\subsection{Interstellar extinction}
We compare our measurement of the extinction with previous ones. In Sec.~\ref{sec:locext}, we measured the extinction from the changes incurred to the slope of the spectrum in the continuum. The extinction found in Brackett $\gamma$ was then converted to $A_{\text{K}_\text{S}}=~3.18 \pm 0.20$ through integration across the $\text{K}_\text{S}$ band, quite stable between our two epochs. Others also estimated the extinction in the direction of GCIRS~7. \citet{1996ApJ...470..864B} obtained $A_{\text{K}}=3.72 \pm 0.13$ via near-infrared (NIR) photometry and an assumed intrinsic color, and later \citet{2003ApJ...597..323B} measured $A_{\text{K}}=3.48 \pm 0.09$. Both values are more than $1\sigma$ higher than our measurement. This discrepancy with archival data could be due to the better spatial filtering offered by GRAVITY thanks to the mono-mode fibers and interferometric spatial resolution, which allows for most of the infrared excess to be rejected from the surrounding material, or to a change in the circumstellar contribution, subject to any variation of the fundamental parameters of the star as it is pulsating \citep{2014A&A...568A..85P}. On the other hand, the interstellar medium (ISM) in the central parsec is known to be clumpy and heterogeneous, responsible for variations of more than one magnitude in the $\text{K}$ band from one line-of-sight to the next and along a given line-of-sight \citep[and references therein]{2016A&A...594A.113C}. The motion of dusty clumps in the ISM over decade-long time scales \citep{2012Natur.481...51G, 2013ApJ...763...78G, 2019A&A...621A..65C, 2020Natur.577..337C} could easily explain variations of the interstellar extinction by a fraction of a $\text{K}$-band magnitude. A value of 3.18 is above average when compared to the extinction map of \citet{2010A&A...511A..18S}; but it is rather typical in view of the extinction map of \citet{2016A&A...594A.113C} who, by using the ISM instead of the stars, was arguably able to prove deeper along the line-of-sight. Therefore, a large part of the excess extinction compared to the field average towards GCIRS~7 is attributable to ISM local to the central parsec, as is possibly  of circumstellar origin in some part.

\begin{figure}
  \centering
    \includegraphics[width=\linewidth]{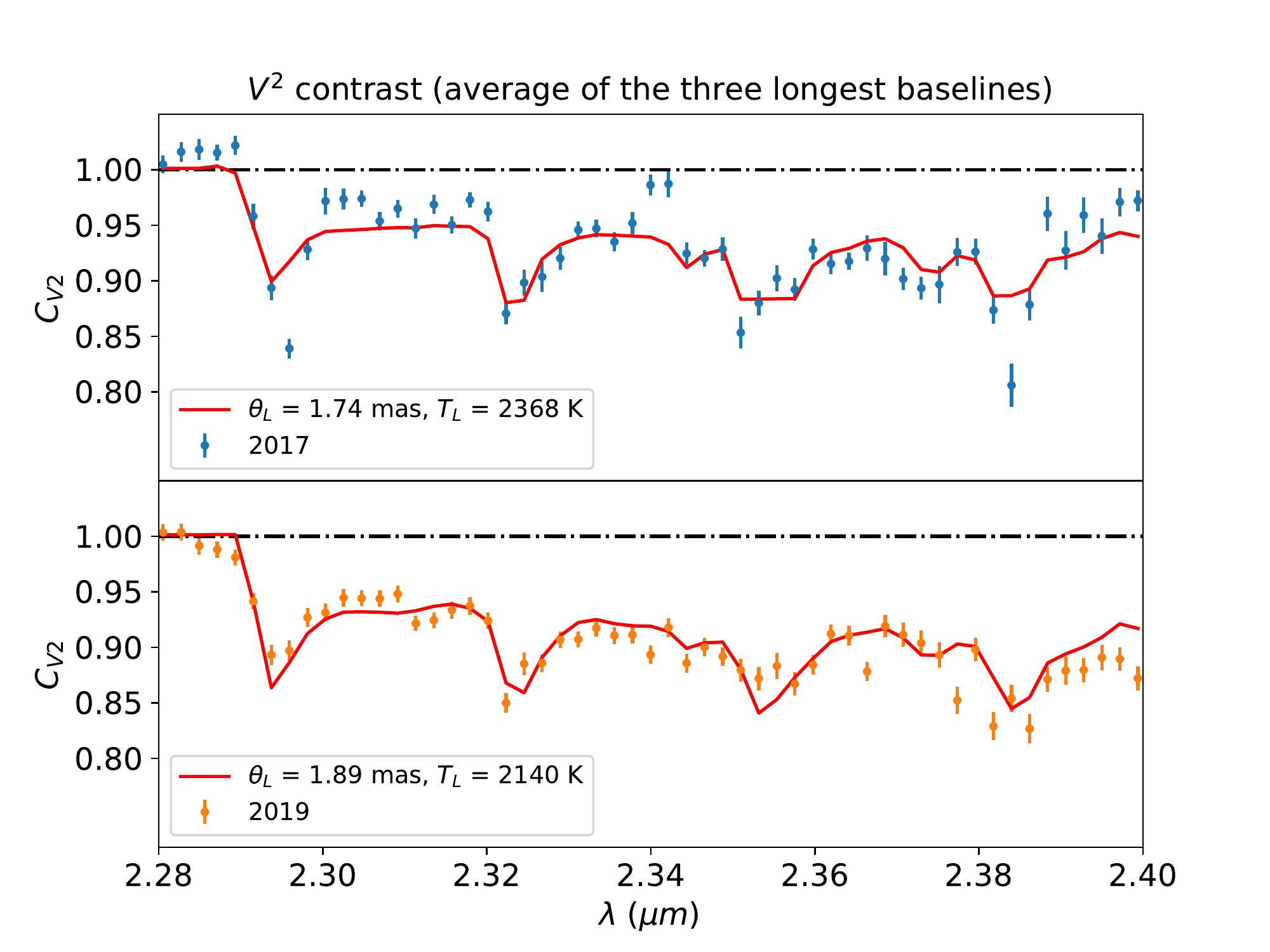}
  \caption{Visibility contrast in the molecular features. The spectral visibility of the three longest baselines have been averaged for each epoch.}\label{fig:Vcon}
\end{figure}

\subsection{Photospheric size}

We measure a photospheric diameter of $\theta^{*}_\text{UD}~=~1.55~\pm~0.03$~mas, yielding $R_*=1368\pm27\,R_\sun$ at $8.246$~kpc \citep{2020A&A...636L...5G}. This value is in good agreement with the values of $\theta^*_\text{UD}=1.5-2$ mas of \citet{2014A&A...568A..85P} on K-band data obtained with AMBER in 2008, but much larger than their H-band measurement on PIONIER 2013 data: $\theta^*_\text{UD}=1.076\pm0.093$ mas. 

\begin{figure}
        \centering  
                \includegraphics[width=\linewidth]{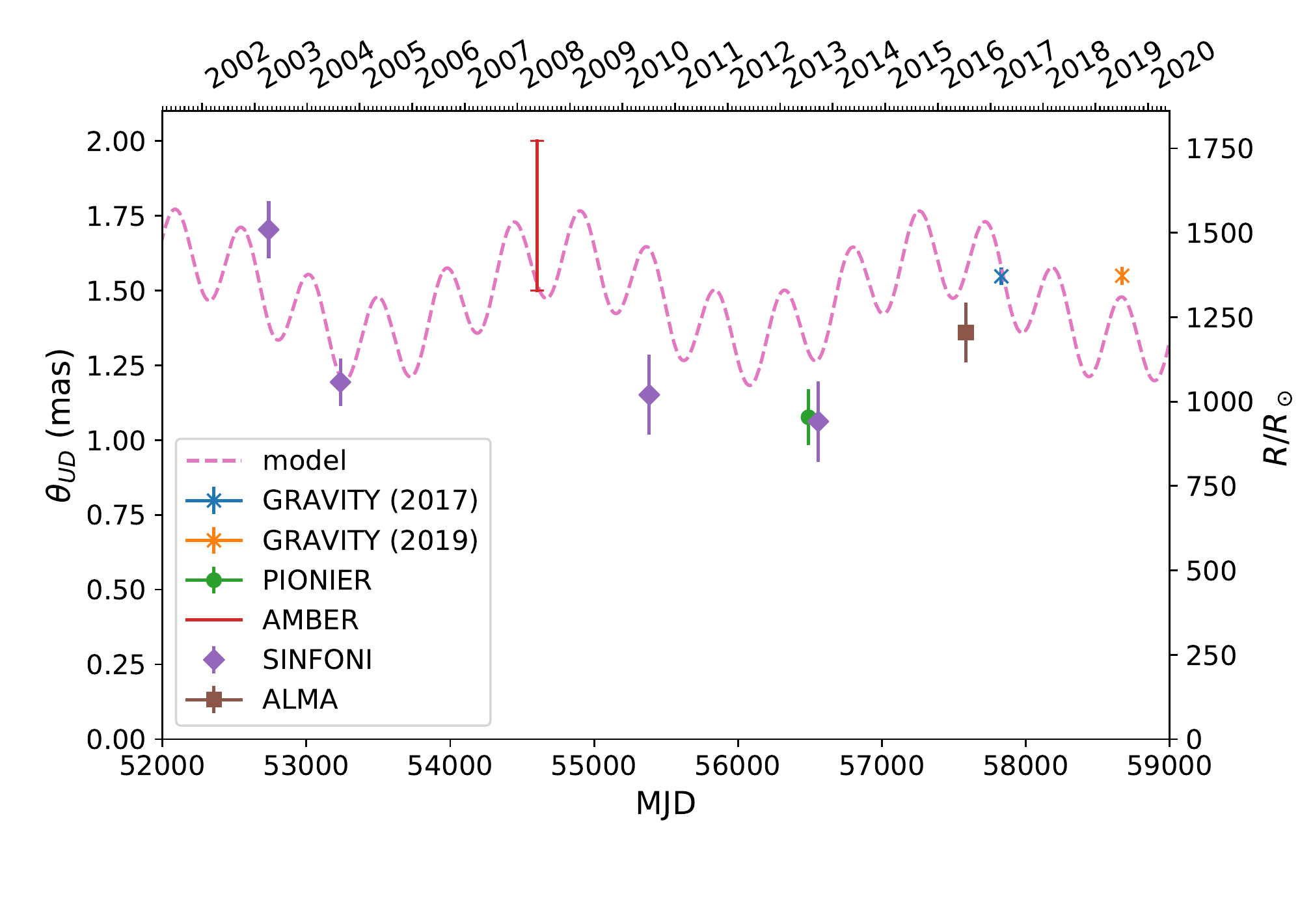}
        \caption{Photospheric uniform-disk and limb-darkened (in the case of SINFONI) diameter estimates from \citet{2014A&A...568A..85P}, \citet{2020PASJ..tmp..162T} and this work. Only the AMBER (K-band), PIONIER (H-band) and GRAVITY (K-band) points are interferometric measurements. The SINFONI (H/K-band) and ALMA (340~GHz) points are spectro-photometric estimates. We provide a sample model using the two periods in \citet{2014A&A...568A..85P}.} 
\label{fig:thetaUD}
\end{figure}

One possible explanation for a smaller photosphere at H-band compared to K-band is the wavelength-dependence of the opacity of negative hydrogen $\mathrm{H}^-$, which determines the opacity of RSGs \citep{gray_2005}. However, this process alone can hardly account for a factor 1.44 in size between the two bands. Therefore, these measurements provide another clue that GCIRS 7 has been pulsating over the last ten years. Such variations in stellar diameter are also corroborated by photometric estimates using SINFONI \citep{2014A&A...568A..85P} and ALMA \citep{2020PASJ..tmp..162T}. The $\theta^*_\text{UD}$ diameters from \citet{2014A&A...568A..85P}, \citet{2020PASJ..tmp..162T} and this work are displayed in Fig.~\ref{fig:thetaUD}, together with a model assuming the pulsation periods (470 and 2620 days) obtained by \citet{2014A&A...568A..85P}. The phases and amplitudes of the two modes of pulsations and the average size were semi-manually adjusted, returning a value for $\chi^2_r$ that is well below $1$. However, given the little data available, too many solutions are possible to make it useful to quote the best-fit parameters. This shows that the data are consistent with pulsations but we are not able to provide further constraints at this stage.

We measure the same photospheric size in 2017 and 2019, with a $3\sigma$ upper limit on the difference of $0.13$~mas ($\simeq8\%$). This is plausibly explained by the observational gap. With the interplay of the two periods (470 and 2620 days) revealed in \citet{2014A&A...568A..85P}, the size of the photosphere could well have varied during the 840 days that separate the observations performed in 2017 and in 2019 and may have come back to a very similar value (Fig.~\ref{fig:thetaUD}). However, it is also possible that the size of the star did not vary between 2017 and 2019 or varied less than expected from the 2013--2017 era. The pulsations of red supergiants are irregular and intertwined with convective mechanisms. Their periods, phases, and amplitudes can vary over short time-scales.

\begin{figure}
  \centering
    \includegraphics[width=\linewidth]{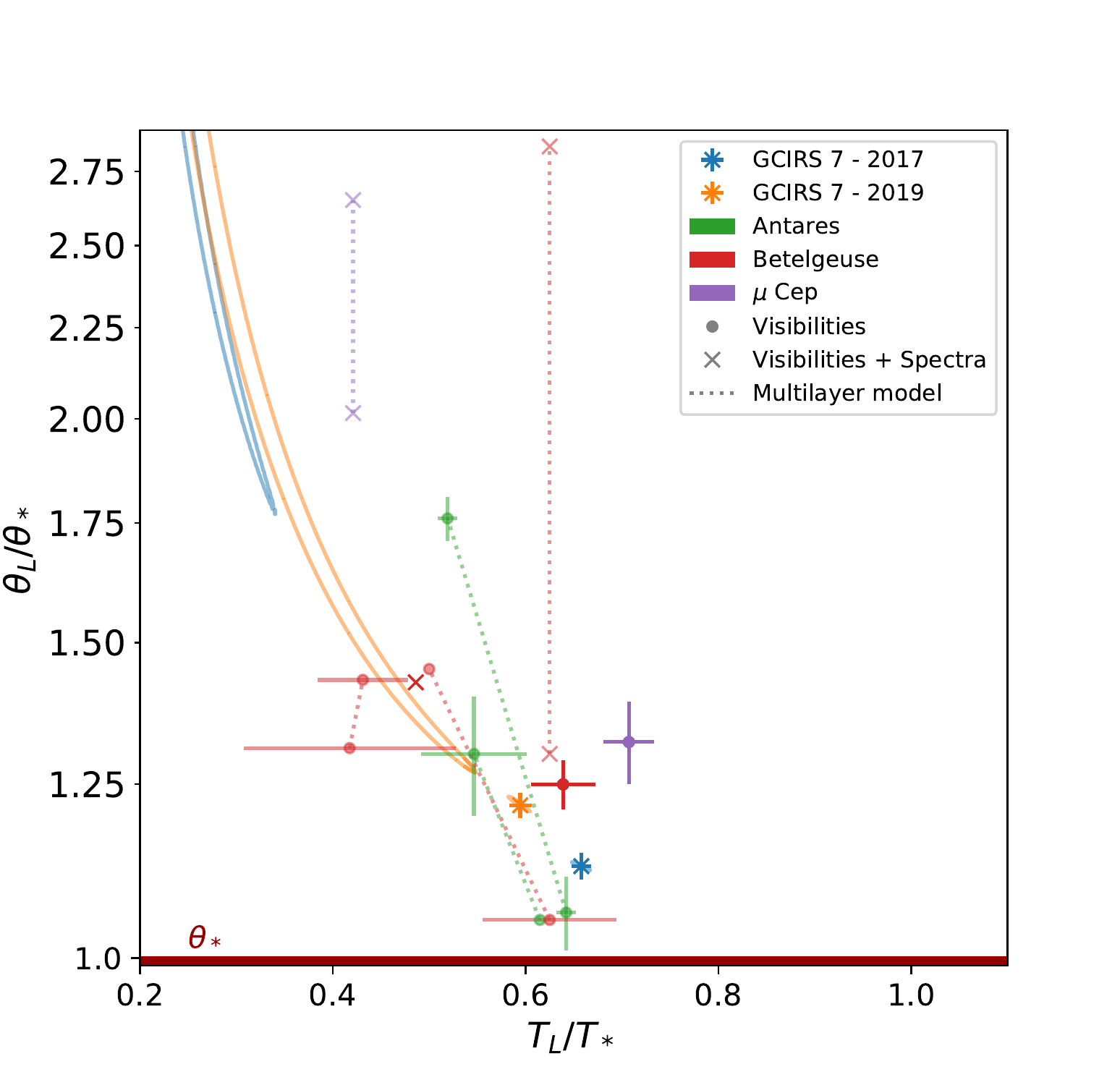} 
  \caption{Physical parameters of the molecular shell of a GCIRS~7 and a sample of stars from the literature derived by single-shell and multiple-shell models by using interferometric and spectroscopic measurements. The contours shown correspond to 1 sigma from their respective hot solutions of GCIRS~7 obtained in this work, which are represented by the crosses with errorbars in blue and orange. The green, red, and purple symbols correspond to previous works for Antares, Betelgeuse, and $\mu$Cep respectively (Table~\ref{tab:allmolspheres}). The circles correspond to models where only visibilities were used, while the crosses represent models which use both visibilities and spectra. The dotted lines connect the inner and the outer layers on the multiple shell models in a same work.}\label{fig:molspherescomp}
\end{figure}

\begin{table}
\caption{MOLsphere models implemented in literature for the closest RSGs.} \label{tab:allmolspheres}
\centering
\begin{tabular}{| c | c | c | c |}
\hline
\multicolumn{4}{|c|}{Single shell models}\\
\hline
$T_*$ (K) & $T_L$ (K) & $\theta_L/\theta_*$ & Reference \\ \hline
3790 & $2680 \pm 100$ & $1.32 \pm 0.07$ & Perrin+ 05 \\
3600  & 1750  & 1.42 & Verhoelst+ 06\\
3600 & $2300\pm 120$  & $1.25 \pm 0.04$ & Montargès+ 14\\ 
3600  & $2362\pm 91$ & $1.16 \pm 0.03$ & This work (2017)\\
3600  & $2182 \pm 80$ & $1.26 \pm 0.03$ & This work (2019) \\
\hline
\multicolumn{4}{|c|}{Multiple shell models (inner and outer layer)}\\
\hline
 \multirow{2}{*}{3800}   & \multirow{2}{*}{1600}  & 2.01 & \multirow{2}{*}{Tsuji 06 ($\mu$Cep)}\\ 
   &  &  2.65 & \\
\hline
\multirow{2}{*}{3600}  & \multirow{2}{*}{2250}  & 1.3 & \multirow{2}{*}{Tsuji 06 ($\alpha$Ori)}\\
  &  & 2.84 & \\
\hline
\multirow{2}{*}{3641} & $1520 \pm 400$ & 1.31 & \multirow{2}{*}{Perrin+ 07}\\
  & $1570 \pm 150$ & 1.43 & \\ \hline
\multirow{2}{*}{3600} & $2250 \pm 250$  & 1.05 & \multirow{2}{*}{Ohnaka+ 09}\\ 
  & 1800  & 1.45 & \\ 
\hline
 \multirow{2}{*}{$3660\pm 120$}   & 2250  & 1.05 & \multirow{2}{*}{Ohnaka+ 13} \\
    & $2000 \pm 200$  & $1.3 \pm 0.1$ & \\ \hline
  \multirow{2}{*}{$3660\pm 120$}  & $2350 \pm 50$ & $1.06\pm 0.05$ & \multirow{2}{*}{Hadjara+ 19}\\
      & $1900 \pm 50$  & $1.76 \pm 0.05$  & \\
\hline
\end{tabular}
\end{table} 

\subsection{The MOLsphere of GCIRS 7 in context}\label{sec:Molspheres}

Here, we compare the results of this work with the parameters of the MOLspheres of Antares, Betelgeuse and $\mu$Cep as derived in various works following similar approaches as ours.

A single thin shell, similar at the one used in this work, was previously modeled in \citet{2005A&A...436..317P} for $\mu$Cep via interferometry in the K band. The work of \citet{2006A&A...447..311V} combined spectroscopy in NIR, MIR, and FIR with interferometry in the K and L bands to characterize the extended atmosphere of Betelgeuse revealing the presence of alumina, while \citet{2006ApJ...637.1040R} presented new MIR spectroscopic data which could not rule out a scenario based only on a cool photosphere ($T=3250$~K). For the same star, the most recent work involving a thin shell model was presented by \citet{2014A&A...572A..17M}, who used K-band interferometry to estimate the temperature, diameter and density of the thin shell.

Several multi-layer models have been also proposed for RSGs. \citet{2006ApJ...645.1448T} used NIR and MIR spectroscopy, together with K-band interferometry, to constrain a two-layer model (inner and outer) for $\mu$Cep and Betelgeuse. A year later, \citet{2007A&A...474..599P} used N-band interferometry to model the close stellar environment of Betelgeuse with the use of another two-layer model estimating the density of the $\text{H}_2$O, SiO, and $\text{Al}_2\text{O}_3$ molecules in the MOLsphere. \citet{2009A&A...503..183O} also resolved spatially the structure of the atmosphere of Betelgeuse and characterized a MOLsphere by using two layers and K-band interferometry. This extended component was also observed by \citet{2011A&A...529A.163O} (K-band interferometry) who found a significant variation of the visibilities and phases in the CO overtone lines between two different epochs, but no significant variation in the continuum. For the star Antares, \citet{2013A&A...555A..24O} used a similar approach (K-band interferometry) characterizing a double layer model at two different epochs, but without any significant variations in temperature or size. A model consisting in several concentric layers was used in \citet{2019MNRAS.489.2595H} to characterize a sample of M stars via K-band interferometry. In the case of Antares, they modeled seven layers between 1.06 and 1.76 stellar radii.

These measurements are listed in Table~\ref{tab:allmolspheres} and displayed in Fig.~\ref{fig:molspherescomp}, as well as the results of this work. The properties of GCIRS~7 are compatible in 2019 within one sigma with all the single shell models, except for \citet{2006A&A...447..311V}, and they are also compatible with the outer shell of Antares from \citet{2013A&A...555A..24O}. On the other hand, 2017 reveals a smaller shell, close to the inner layer of Antares from \citet{2019MNRAS.489.2595H}. 

In addition, considering the multiple shell models, both the inner and the outer shell of \citet{2007A&A...474..599P} and the outer shell of \citet{2013A&A...555A..24O} are also compatible within 1$\sigma$ to the through pointing to the cold solution in 2019. This is an indication of the shortcomings of the single shell model to interpret 2019 data for GCIRS~7. 

\subsection{Size and expansion of the molecular shell}

The ratio of the shell diameter to the photospheric diameter ($1.16$--$1.26$) is very similar to previous findings. \citet{2005A&A...436..317P} found a ratio of 1.32 for $\mu$ Cep, \citet{2014A&A...572A..17M} found 1.25 for Betelgeuse. This is still compatible with the lower altitude layers of $\alpha$~Sco for which \citet{2019MNRAS.489.2595H}
found ratios in the range 1.06 to 1.76, and temperatures between 2360 and 1900$\,$ K using a multi-layer model (see the next section for the discussion of the temperature). In addition, \citet{2017A&A...606L...1W} measured a MOLsphere extending farther than 2.5 times the photosphere for V766 Cen. For the same star, \citet{2017A&A...597A...9W} measured a ratio 1.49 between the photospheric radius and the radius of a layer of \ion{Na}{i} causing a line in emission at 2.205 $\mu\mathrm{m}$.

The uniform disk diameters for the molecular band presented on Fig.~\ref{fig:FnUD1719} show globally higher values in 2019 than in 2017. This is also evidenced by the significantly lower visibility contrast in 2019 beyond 2.33$\,\mu$m (Fig.~\ref{fig:Vcon}). This is confirmed by the analysis with the single-layer model where the hot solutions found in 2017 and 2019 in Fig. \ref{fig:chimap} are different by 4$\sigma$. Our results are compatible with a layer expansion of 8\% leading to a layer temperature decrease from 2017 to 2019. 

However, the single shell model constitutes a simplified approach for a MOLsphere. Indeed, although a strong contribution of the shell due to a larger size or a lower temperature is found in 2019, the apparent difference in the shell between the two epochs may be a consequence of the degeneracies of the parameters. A more complex model would be needed to solve the degeneracies and, therefore, to be able to conclude about a possible shell expansion.

\subsection{Column density of CO}
The optical depth profiles shown in Fig.~\ref{fig:tausch} are consistent in 2017 and 2019 except for the peaks where the model saturates. With these optical depth profiles, we used the method of \citet{1994ApJS...95..535G} to estimate the column density of CO. The method uses the first band head of CO at 2.293\,$\mu$m and the temperature of the shell $T_L$ measured for each epoch. Using the parameter of the hot solution and assigning the optical depth measured with the model in the 2.293$\,\mu$m channel to the first band head of the CO overtone yields, for each epoch we have:

\begin{itemize}
    \item $N_{\mathrm{CO}}(2017)=(2.68\pm 0.38) \times 10^{20}$\,mol\,cm$^{-2}$;
    \item $N_{\mathrm{CO}}(2019)=(2.15 \pm 0.26) \times 10^{20}$\,mol\,cm$^{-2}$.
\end{itemize}

These values are similar to what \citet{2006ApJ...645.1448T} measured for $\mu$Cep and Betelgeuse: $N_{\mathrm{CO}}\approx 10^{20}$\,mol\,cm$^{-2}$. Also for Betelgeuse, by using several isotopes \citet{2014A&A...572A..17M} measured a higher density: $N_{\mathrm{CO}}=3.01^{+2.0}_{-0.5} \times 10^{21}$\,mol\,cm$^{-2}$. Relative to Antares, our densities are compatible with the value measured by \citet{2013A&A...555A..24O}: $N_{\mathrm{CO}}\approx 10^{19-20}$\,mol\,cm$^{-2}$ at 1.3 $R_*$, and in the range of values found by \citet{2019MNRAS.489.2595H} ($10^{21.5}-10^{19.2} \,\mathrm{mol}\,\mathrm{cm}^{-2}$).

The optical depths of Fig.~\ref{fig:tausch} reveal values slightly larger for 2017 than for 2019 as long as the wavelength increases, except for the first head band, which is the same at the two epochs. This is the effect of the monotonic increase of the diameter mentioned in the uniform disk approach (Fig.~\ref{fig:FnUD1719}). In addition, in 2017 none the head bands except for the first one can be reproduced by the optical depths of the thin shell due to saturation, while the first two band heads are reproduced in 2019. This is an indicator that the density of the MOLsphere of GCIRS~7 might have decreased.

It is possible to test if this decrease can be due to an expansion of the molecular shell by scaling the measured density by the ratio of the surface of the shell at the two epochs. Assuming no sudden mass loss episode, if the shell has expanded since 2017 at the measured diameter (from 1.74 to 1.89 mas), the column density of CO in 2019 would be $N_\text{CO}~=~2.27\times~10^{20}$\,mol\,cm$^{-2}$, which is 1$\sigma$ compatible with the measured value. Therefore, our single-shell interpretation allows for a possible expansion of the molecular envelope, but the fact that the model is simplified does not allow for it to be fully confirmed without a deeper study involving an extended atmosphere approach. 

Indeed, although CO is the main component of the molecular shell, other molecular species such as water have also been observed in the MOLspheres of other RSGs \citep{2006A&A...447..311V, 2006ApJ...637.1040R, 2007A&A...474..599P, 2014A&A...572A..17M}, and the decrease of the temperature of the molecular shell may also play a role in the density variation. For GCIRS~7, the molecular shell in 2017 could have been warm enough for water to dissociate ($T=2400$~K), although it may still have existed at $T=2200$~K in 2019. 

The location of GCIRS~7 is particular as it is located at less than 1 pc from a supermassive black hole and plenty of massive young stars \citep{1995ApJ...447L..95K} and it is known that the most external layers are being blown away by the effect of their wind \citep{2020PASJ..tmp..162T, 1991ApJ...371L..59Y, 1991ApJ...378..557S}. The question of whether or not their presence affects the observed molecular shell of GCIRS~7 could be addressed by the use of a more complex model to fully characterize the outer atmosphere of GCIRS 7, such as that of \citet{2019MNRAS.489.2595H}. In addition, either increasing the signal to noise ratio in the K band or obtaining more interferometric data at longer baselines, as well as exploring other wavelengths (H, L bands), would bring more constraints to better characterize the outer atmosphere of GCIRS 7. Such approaches would help to understand the nature of the molecular shell observed, as well as its temporal evolution.

\section{Conclusion}\label{sec:conc}

We report on the spectro-interferometry of GCIRS~7 in the K band using GRAVITY at ESO/VLTI. With the sensitivity of current interferometers, we proved that wavelength-dependent structures can be observed in evolved stars even if they are not fully resolved. 

We find that GCIRS~7 presents the behavior of a typical RSG. We detect a molecular shell above the photosphere and estimate the sizes and temperatures for the photosphere and the shell as well as the column density for CO, based on optical depths constrained with a single-layer model. This is, to date, the RSG star with the smallest apparent size and the farthest for which a molecular shell has been spatially resolved from the star and characterized. We also obtained an estimate of the local interstellar extinction with the spectral data.

The extinction ($A_{\text{K}_\text{S}}=3.18\pm0.16$) and size of the photosphere ($\theta^*_\text{UD}=1.55\pm0.03$) were the same within uncertainties at the two epochs in 2017 and 2019.
However, the photospheric size must have changed from $>1.5$~mas in 2008 to $1.1\pm0.1$~mas in 2013 \citep{2014A&A...568A..85P} and back to $1.55\pm0.03$~mas in 2017--2019.

The spectro-differential visibility signal demonstrate the presence of CO above the photosphere. In the context of a thin spherical shell model, the temperature ($\approx2200$--$2400$~K) and diameter ($10$--$20\%$ larger than the photosphere) of this shell are in line with what has been found for other similar RSG stars. The size and temperature of the shell have significantly changed between the two epochs and are compatible with an expansion.

The column density for the molecular shell presents a value in the same line than the column density of the shells for other RSGs measured with similar methods. The model fails to reproduce all the band heads in 2017 except the first, while in 2019 the first two band heads are reproduced. This suggests that the density must have been higher in 2017. An interpretation based on a shell expansion from 2017 to 2019 is compatible with our data.

This work corresponds to a first-order description of the outer atmosphere of GCIRS~7. Overall, our results support the interpretation in terms of stellar pulsations proposed by \citet{2014A&A...568A..85P} and hint at an expansion of a molecular shell. Follow-up observations over a good fraction of the $\approx 2800$-day period with contemporaneous H and K photometry, spectroscopic effective temperature, and interferometric size measurements, together with a more detailed multi-layer model made relevant by the spectral resolution of GRAVITY, would be needed to further confirm the pulsations and study the associated mass-loss processes.

\begin{acknowledgements} This research has made use of the Jean-Marie Mariotti Centre \texttt{LITpro} service co-developed by CRAL, IPAG and Lagrange, \footnote{LITpro software available at http://www.jmmc.fr/litpro} \texttt{OifitsExplorer} service \footnote{Available at http://www.jmmc.fr/oifitsExplorer}, \texttt{Aspro2} service \footnote{Available at http://www.jmmc.fr/aspro2}, , \texttt{SearchCal} service \footnote{Available at http://www.jmmc.fr/searchcal} co-developed by LAGRANGE and IPAG, and of CDS Astronomical Databases SIMBAD and VIZIER \footnote{Available at http://cdsweb.u-strasbg.fr/}. This publication makes use of data products from the Two Micron All Sky Survey, which is a joint project of the University of Massachusetts and the Infrared Processing and Analysis Center/California Institute of Technology, funded by the National Aeronautics and Space Administration and the National Science Foundation. A.A. and P.G. were supported by Funda\c{c}\~{a}o para a Ci\^{e}ncia e a Tecnologia, with grants reference UIDB/00099/2020 and SFRH/BSAB/142940/2018. The corresponding author would also like to thank all the support astronomers, the operations team, the logistics and service staff of Paranal Observatory for their help during the observations. SvF, FW \& AJ-R acknowledge support by the MaxPlanck International Research School. 
\end{acknowledgements}

\bibliographystyle{aa} 
\bibliography{main.bib}

\end{document}